%% file: cr_main.tex
\documentclass[sigconf, screen]{acmart}

\usepackage{fancyhdr}

\usepackage{algorithmic}
\usepackage{algorithm}
\usepackage{graphicx}
\usepackage{textcomp}
\usepackage{xcolor}
\usepackage{balance}
\usepackage{tablefootnote}
\usepackage{multirow}
\usepackage{pifont}
\usepackage{soul}
\usepackage{etoolbox}
\usepackage{stmaryrd}

\usepackage{booktabs}
\usepackage[utf8]{inputenc}
\usepackage{array}
\usepackage{makecell}

\usepackage{colortbl}

\usepackage{array}
\newcolumntype{x}[1]{>{\centering\arraybackslash\hspace{0pt}}p{#1}}

\newcommand{\figref}[1]{Figure~\ref{#1}}
\newcommand{\tabref}[1]{Table~\ref{#1}}
\newcommand{\secref}[1]{Section~\ref{#1}}
\newcommand{\algoref}[1]{Algorithm~\ref{#1}}

\newcommand{\mmhf}{\texttt{MMH4}\xspace}
\newcommand{\hacc}{\texttt{HACC}\xspace}
\newcommand{\mtag}{$\mathrm{TAG}$\xspace}

\usepackage{xspace}

\AtBeginDocument{%
  }

\newcommand*\circled[1]{\tikz[baseline=(char.base)]{
            \node[shape=circle,fill,inner sep=1pt] (char) {\textcolor{white}{#1}};}}

\usepackage[firstpage=true,color=black,angle=90,scale=2.0,position={9.08,-7.0}]{background}
\SetBgContents{Accepted Paper: ISCA 2024, Argentina}

\acmBadgeR[https://www.acm.org/publications/policies/artifact-review-and-badging-current\#available]{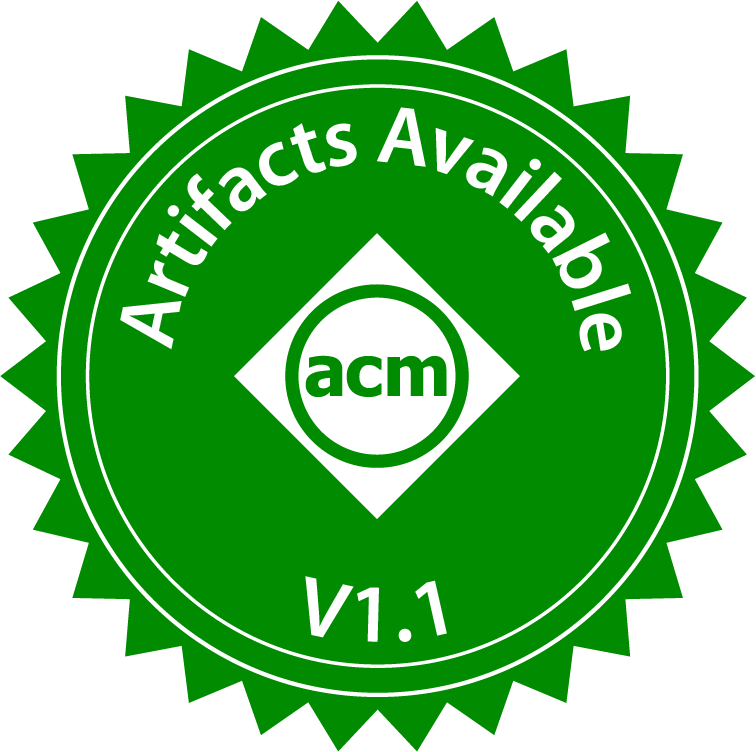}
\acmBadgeR[https://www.acm.org/publications/policies/artifact-review-and-badging-current\#functional]{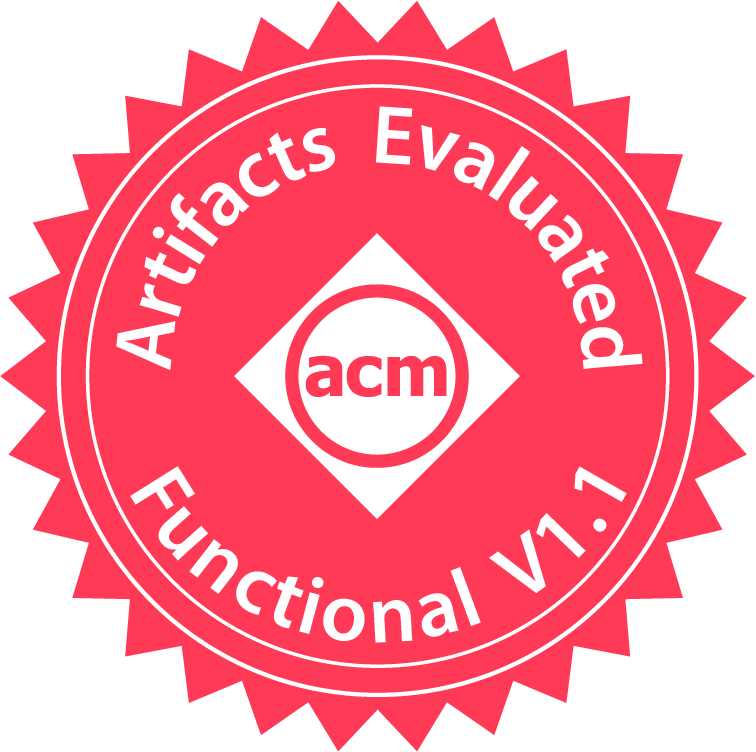}

\input{util/acm_copyrights}

\begin{document}

\title{NeuraChip: Accelerating GNN Computations with a Hash-based Decoupled Spatial Accelerator}

\input{util/authors}

\input{sections/00_abstract}

\ccsdesc[500]{Computer systems organization~Multicore architectures}
\ccsdesc[300]{Computer systems organization~Interconnection architectures}
\ccsdesc[100]{Computing methodologies~Neural networks}
\ccsdesc[300]{Theory of computation~Graph algorithms analysis}
\ccsdesc[100]{Hardware~Hardware accelerators}

\keywords{Graph Neural Networks (GNN),
Decoupled Computations,
Spatial Accelerators,
Sparse Matrix Multiplication (SpGEMM),
On-chip Memory, Hardware-software co-design}




\def\authors{Shivdikar et al}



\maketitle


\input{sections/01_introduction}

\input{sections/02_background}

\input{sections/03_neurachip}

\input{sections/04_design_space}

\input{sections/05_evaluation}

\input{sections/08_related_work}

\input{sections/09_conclusion}

\input{sections/99_artifact}

\input{sections/acknowledgement}

\balance
\bibliographystyle{ACM-Reference-Format}
\bibliography{bibliography/biblio}

\end{document}

%% file: util/authors.tex
\author{
    Kaustubh~Shivdikar$^1$ \hspace{0.1em}
	Nicolas~Bohm~Agostini$^1$ \hspace{0.1em}
	Malith~Jayaweera$^1$ \hspace{0.1em}
	Gilbert~Jonatan$^2$ \hspace{0.1em}
        Jos\'e~L.~Abell\'an$^3$ \hspace{0.1em}
        Ajay~Joshi$^4$ \hspace{0.1em}
        John~Kim$^2$ \hspace{0.1em}
        David~Kaeli$^1$
}

\affiliation{\vspace{0.1in} \institution{
        $^1$Northeastern University \hspace{0.3em}
	$^2$KAIST \hspace{0.3em}
 	$^3$Universidad de Murcia
        $^4$Boston University \hspace{0.3em}
 }
 \country{ \{shivdikar.k, bohmagostini.n, malithjayaweera.d, d.kaeli\}@northeastern.edu \\ \{gilbertjonatan, jjk12\}@kaist.ac.kr, joshi@bu.edu, jlabellan@um.es
} 
}

\renewcommand{\shortauthors}{Shivdikar et al.}

%% file: sections/00_abstract.tex
                                          
\begin{abstract}

Graph Neural Networks (GNNs) are emerging as a formidable tool for processing non-euclidean data across various domains, ranging from social network analysis to bioinformatics.
Despite their effectiveness, their adoption has not been pervasive because of scalability challenges associated with large-scale graph datasets, particularly when leveraging message passing.
They exhibit irregular sparsity patterns, resulting in unbalanced compute resource utilization.
Prior accelerators investigating Gustavson's technique adopted look-ahead buffers for prefetching data, aiming to prevent compute stalls.
However, these solutions lead to inefficient use of the on-chip memory, leading to redundant data residing in cache.

To tackle these challenges, we introduce NeuraChip, a novel GNN spatial accelerator based on Gustavson's algorithm. 
NeuraChip decouples the multiplication and addition computations in sparse matrix multiplication.
This separation allows for independent exploitation of their unique data dependencies, facilitating efficient resource allocation.
We introduce a rolling eviction strategy to mitigate data idling in on-chip memory as well as address the prevalent issue of memory bloat in sparse graph computations.
Furthermore, the compute resource load balancing is achieved through a dynamic reseeding hash-based mapping, ensuring uniform utilization of computing resources agnostic of sparsity patterns.
Finally, we present NeuraSim, an open-source, cycle-accurate, multi-threaded, modular simulator for comprehensive performance analysis.

Overall, NeuraChip presents a significant improvement, yielding an average speedup of $22.1\times$ over Intel's MKL, $17.1\times$ over NVIDIA's cuSPARSE, $16.7\times$ over AMD's hipSPARSE, and $1.5\times$ over prior state-of-the-art SpGEMM accelerator and $1.3\times$ over GNN accelerator.
The source code for our open-sourced simulator and performance visualizer is publicly accessible on GitHub\footnotemark.

\vspace{-0.5em}

\end{abstract}

%% file: sections/01_introduction.tex


\vspace{-1.8em}

\begin{figure}[htbp]
	\centering
	\includegraphics[width=0.48\textwidth]{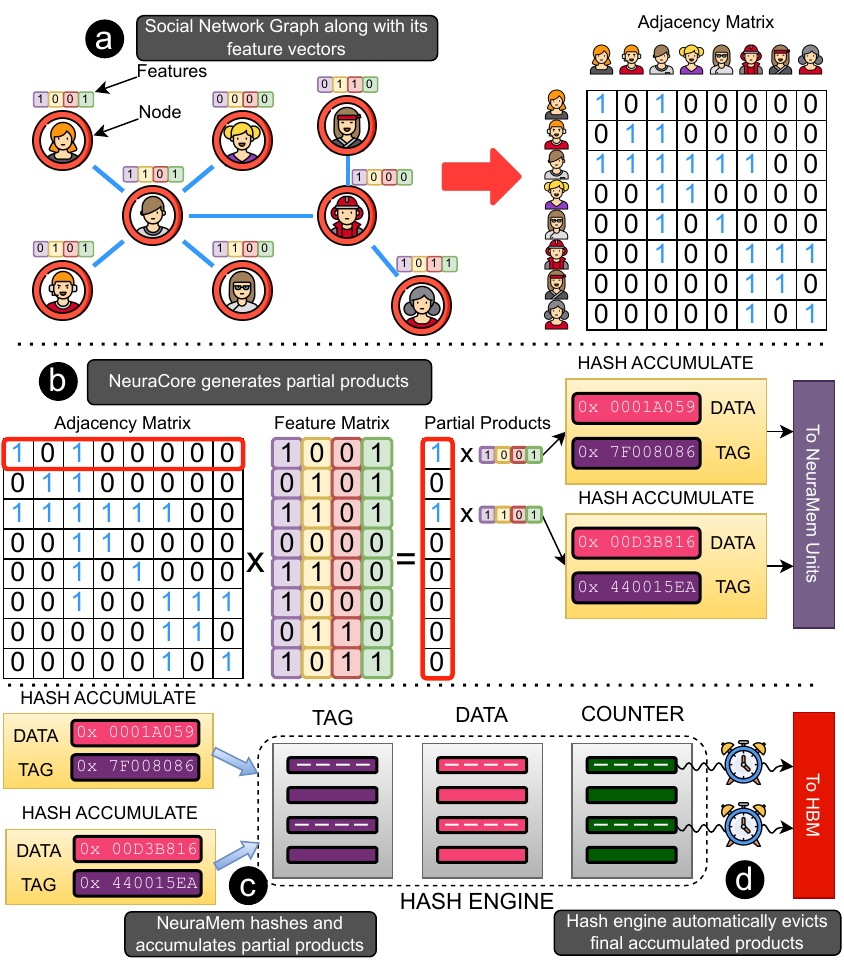}
	\vspace{-2.0em}
	\caption{NeuraChip overview: (a) Aggregation phase of GCN, (b) NeuraCore generates partial products, (c) NeuraMem accumulates partial products, (d) writes back to HBM.}
	\label{fig:teaser}
\vspace{-1.4em}
\end{figure}

\section{Introduction}

\footnotetext{\url{https://github.com/NeuraChip/neurachip}}

\begin{figure*}[tbp]
	\centering
	\includegraphics[width=0.99\textwidth]{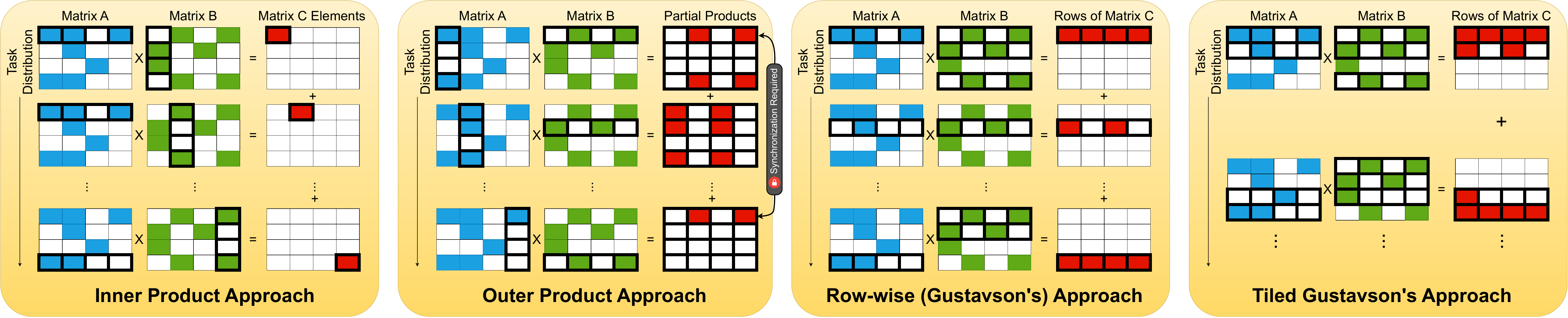}
	\caption{Matrix multiplication approaches, each showcasing varying degrees of data reuse for input and output matrices.}
	\label{fig:matmul_types}
\vspace{-1.0em}
\end{figure*}

Deep Neural Networks have proven to be powerful model for solving problems that rely on data with an underlying Euclidean or grid-like structure, such as computer vision, natural language processing, and audio vision.
In contrast, Graph Neural Networks (GNNs) have emerged as a powerful tool for handling non-Euclidean data (e.g., social networks on the scale of billions~\cite{smith_2020}), achieving impressive performance across various domains such as social science, chemistry, and bioinformatics~\cite{zhou2020graph, xu2018powerful}.
However, the computational complexity of GNNs, especially when working with ultra-sparse, large-scale graph datasets, poses challenges due to architectural limitations of traditional hardware (i.e., CPUs / GPUs)~\cite{shivdikar2023enabling}.
Moreover, GNNs predominantly adopt a recursive neighborhood aggregation (i.e., message passing) methodology, in which each node aggregates the feature vectors of its neighboring nodes to derive its own updated feature vector.
The scalability of message passing in GNNs, when applied to large graph structures, poses a significant bottleneck, especially as the size of the graphs surpasses the capacity of on-chip memory in today's CPUs and GPUs~\cite{jonatan2024scalability}. 
This leads to redundant and time-consuming memory transactions to fetch data from the main memory~\cite{baruah2021gnnmark}.
We identify three main bottlenecks causing redundant memory transactions and propose corresponding hardware/software solutions.

\textbf{Diverse Data Dependence Patterns:} The process of neighborhood aggregation in graphs can be split into a multiplication stage, followed by a merge/reduction stage.
The multiplication stage creates partial products by multiplying the adjacency matrix of the input graph with the feature matrix.
The reduction stage accumulates (merges) the partial products to update the node feature vectors.
The multiplication stage's operands depend on data stored in the high-bandwidth memory (HBM). 
In contrast, the reduction stage's operands depend on data located within the on-chip memory.
Utilizing a singular computational resource for both multiplication and accumulation operations proves suboptimal, as mapping multiplication operations on computing resources tends to compromise the efficiency of mapping the accumulation operations (due to varying data dependency patterns).
To effectively address these distinct data dependencies during these stages, we present two dedicated components in our custom accelerator, tailored to specific data dependencies: NeuraCore, which executes multiplication tasks, and NeuraMem, which accumulates the intermediate partial products generated by NeuraCore (\figref{fig:teaser}).

\textbf{Uneven Hardware Resource Utilization:} The multiplication and accumulation stages are characterized by distinct architectural implications.
The multiplication stage typically stalls due to data starvation (unavailability of input graph and feature matrix elements), whereas the accumulation stage suffers from uneven partial product distribution due to irregular sparsity patterns. 
To address these issues, we present two strategies, each catering to their respective problems.
(a) \textit{Multiplication mapping:} We implement a \textit{tiled row-wise product approach} to partition the computation into distinct tasks, which are then dynamically allocated to NeuraCore, depending on its utilization.
The row-wise product method, also known as Gustavson's algorithm, is a popular choice among recent accelerators such as Gamma~\cite{zhang2021gamma}, MatRaptor~\cite{srivastava2020matraptor}, and SPADA~\cite{li2023spada} as this approach has shown high-efficiency when targeting sparse matrix computations in the aggregation stages of GNNs.
Developing dedicated components for multiplication enables mapping these operations to NeuraCore, independent of the accumulation stage, thus leveraging the locality of the input data. 
(b) \textit{Accumulation Mapping:} We present a dynamic reseed hash-based mapping agnostic to sparsity patterns.
This allows us to evenly distribute the partial products among the NeuraMem accumulation units.

\input{tables/bloat}

\textbf{Memory Bloat:} Incorporating the row-wise product approach enhances input data locality but creates a large number of partial products~\cite{baek2021innersp}.  Table~\ref{tab:bloat} presents memory bloat for SpGEMM workload across various sparse graph datasets.

\begin{equation}
\label{eq:bloat}
\mathrm{Bloat\ Percent} = \frac{pp_{\mathrm{interim}} - nnz_{\mathrm{output}}}{nnz_\mathrm{output}} \times 100
\end{equation}

\noindent
We define bloat percent as shown in Equation~\ref{eq:bloat}, wherein $pp_{\mathrm{interim}}$ denotes the count of intermediate partial products and $nnz_{\mathrm{output}}$ signifies the count of non-zero elements in the resultant product matrix.
Although tiling the computation partially mitigates this issue, it does not fully resolve it.
Prior solutions like Gamma~\cite{zhang2021gamma} have relied on large explicitly managed cache systems like FiberCache, which consumes up to $72\%$ of the total chip's area.
To address the memory bloat issue, we present a rolling eviction strategy, which automatically evicts a partial product from the on-chip memory once all contributing partial products have been fully accumulated. 
We enable this using an eviction counter integrated with the on-chip memory hashtables.

Our complete accelerator design is named NeuraChip, a spatial accelerator featuring Coarse-Grained Reconfigurable Array (CGRA) on-chip interconnects. 
NeuraChip is versatile due to its ability to efficiently compute graphs with varying degrees of sparsity and sparsity patterns.

\noindent
In this paper, we make the following contributions: 
\begin{itemize}
    \item \textbf{Heterogeneous Processing Approach:} We present NeuraChip, an innovative GNN spatial accelerator featuring a decoupled computation pipeline~\cite{smith1982decoupled}. Segregating multiplication and accumulation operations into dedicated components, we optimize data reuse through strategic mapping.

    \item \textbf{Adaptive Hash-Based Compute Mapping:} Our approach introduces a flexible, dynamic reseeding hash-based compute mapping (DRHM) tailored for GNN workloads. DRHM benefits from the constant lookup times characteristic of hash functions, while also mapping tasks evenly across all computing resources by generating a new seed at predetermined intervals of computation.
    This approach achieves a uniform workload distribution independent of the sparsity patterns in graph workloads.
    
    \item \textbf{Mechanism for Rolling Evictions:} We propose a rolling eviction strategy to combat the issue of memory bloat. Enhanced on-chip hash tables enable the removal of partial products, effectively reducing memory congestion caused by their generation.
    
\end{itemize}

NeuraChip excels when compared to Intel MKL running on an Xeon CPU, surpassing it by a factor of $22.1\times$. Additionally, when compared against NVIDIA's H100 GPU using the CUSP library, NeuraChip achieves a performance boost of $13.3\times$. In comparison to the prior leading sparse matrix multiplication accelerator, Gamma, NeuraChip showcases an average performance improvement of $1.5\times$. Furthermore, NeuraChip outperforms the state-of-the-art GNN accelerator, FlowGNN~\cite{sarkar2023flowgnn}, by an average factor of $1.3\times$.

%% file: tables/bloat.tex
\begin{table}
\centering
\caption{SpGEMM memory bloat analysis across various hyper-sparse graph datasets}
\vspace*{-3mm}
\label{tab:bloat}
\begin{tabular}{l | x{0.5in} x{0.495in} x{0.495in} x{0.495in}}
\arrayrulecolor{black}\toprule
   \textbf{Dataset} & \textbf{Node Count} & \textbf{Edge Count} & \textbf{Sparsity (\%)} & \textbf{Bloat Percent} \\ [0.5ex] 
\arrayrulecolor{black}\toprule
\arrayrulecolor{black}\toprule

2cubes\_sphere & $101492$ & $1647264$ & $99.9840$ & $205.87$ \\
\arrayrulecolor{black!20}\midrule
ca-CondMat & $23133$ & $186936$ & $99.9651$ & $75.23$ \\
\arrayrulecolor{black!20}\midrule
cit-Patents & $3774768$ & $16518948$ & $99.9999$ & $19.32$ \\
\arrayrulecolor{black!20}\midrule
email-Enron & $36692$ & $367662$ & $99.9727$ & $68.90$ \\
\arrayrulecolor{black!20}\midrule
filter3D & $106437$ & $2707179$ & $99.9761$ & $326.34$ \\
\arrayrulecolor{black!20}\midrule
mario002 & $389874$ & $2101242$ & $99.9986$ & $99.43$ \\
\arrayrulecolor{black!20}\midrule
p2p-Gnutella31 & $62586$ & $147892$ & $99.9962$ & $10.21$ \\
\arrayrulecolor{black!20}\midrule
poisson3Da & $13514$ & $352762$ & $99.8068$ & $297.92$ \\
\arrayrulecolor{black!20}\midrule
scircuit & $170998$ & $958936$ & $99.9967$ & $66.13$ \\
\arrayrulecolor{black!20}\midrule
web-Google & $916428$ & $5105039$ & $99.9994$ & $104.27$ \\
\arrayrulecolor{black!20}\midrule
amazon0312 & $400727$ & $3200440$ & $99.9980$ & $97.21$ \\
\arrayrulecolor{black!20}\midrule
cage12 & $130228$ & $2032536$ & $99.9880$ & $127.23$ \\
\arrayrulecolor{black!20}\midrule
cop20k\_A & $121192$ & $2624331$ & $99.9821$ & $327.07$ \\
\arrayrulecolor{black!20}\midrule
facebook & $4039$ & $60050$ & $99.1519$ & $2872.80$ \\
\arrayrulecolor{black!20}\midrule
m133-b3 & $200200$ & $800800$ & $99.9980$ & $26.93$ \\
\arrayrulecolor{black!20}\midrule
offshore & $259789$ & $4242673$ & $99.9937$ & $205.45$ \\
\arrayrulecolor{black!20}\midrule
patents\_main & $240547$ & $560943$ & $99.9990$ & $14.18$ \\
\arrayrulecolor{black!20}\midrule
roadNet-CA & $1971281$ & $5533214$ & $99.9999$ & $35.75$ \\
\arrayrulecolor{black!20}\midrule
webbase-1M & $1000005$ & $3105536$ & $99.9997$ & $36.02$ \\
\arrayrulecolor{black!20}\midrule
wiki-Vote & $8297$ & $103689$ & $99.8494$ & $148.09$ \\
     \arrayrulecolor{black}\bottomrule
\end{tabular}



\vspace{-1.9em}
\end{table}

%% file: sections/02_background.tex
\section{Background and Motivation}

\subsection{Fundamentals of GNN Workloads}

Graph neural networks (GNNs) are capable of extracting important features such as structural motifs (i.e., arrangements of nodes, edges, and metadata)  by learning not just the individual characteristics of each element (i.e., a node in the graph), but also the interconnections (i.e., the interrelationships between nodes) between elements. 
They utilize convolution operations to extract various features from the graph. 
The methodology employed is neighborhood aggregation, where the final feature vector for each vertex is computed by iteratively aggregating and transforming the input feature vectors of adjacent vertices. 
This process includes two steps, which are called the {\em aggregation and combination stages}. 
This process is carried out iteratively, and after $k$ iterations through these stages, the resultant feature vector of the target vertex signifies the distinct structural data of the vertex's $k$-hop vicinity.

For instance, a Graph Convolutional Network (GCN) is one such GNN model. 
Equation~\ref{eq:gcn} below computes the forward propagation for a single layer in a GCN.
\begin{equation}
\label{eq:gcn}
    X^{(l+1)} = \sigma(AX^{(l)}W^{(l)})
\end{equation}
where $A$ represents the adjacency matrix of the graph, where each row lists the interconnections of a vertex to all other vertices in the graph. $X^{(l)}$ refers to the input feature vectors of every vertex in the $l^{th}$ layer of matrix $X$. 
$W$ contains the GNN's model parameters, which are obtained through model training. $\sigma()$ represents the non-linear activation functions like ReLU (Rectified Linear Unit).

\begin{figure}[tbp]
	\centering
	\includegraphics[width=0.48\textwidth]{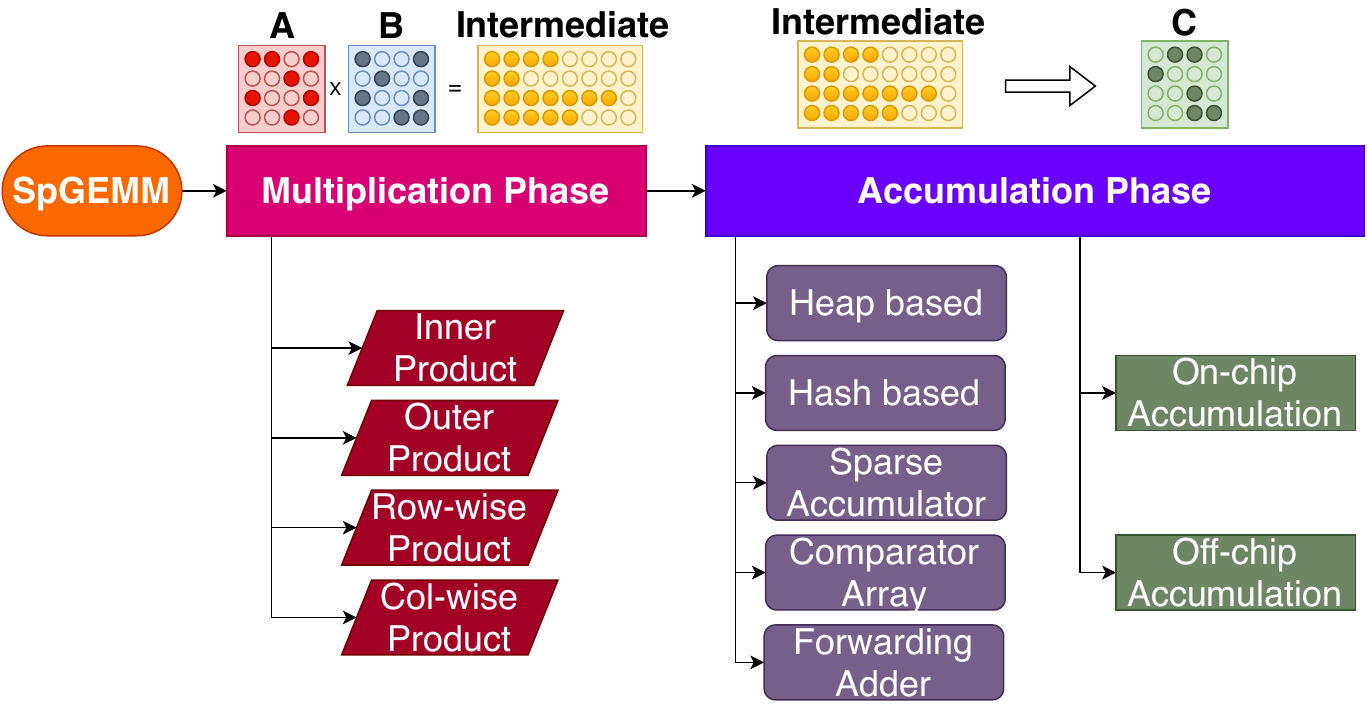}
	\caption{Multiplication and Accumulation phase techniques.}
	\label{fig:spgemm_stages}
\vspace{-1.0em}
\end{figure}

\subsection{Architectural Implications of GNNs}

\textbf{Aggregation Stage}: 
The aggregation stage in GNN workloads is critical for capturing the structural information of graphs. It involves gathering and summarizing information from a node's neighbors, which can be a challenging task given the irregular data structures common in graph-based data.
This is typically computed with sparse matrix multiplication kernels.
Given the high level of sparsity in input graphs, typically above $99\%$, this stage is characterized by random access patterns in memory, which presents a challenge for traditional architectures that are more suited for linear data access. Additionally, the irregular sparsity patterns often lead to workload imbalance on computing resources~\cite{shivdikar2022accelerating}, which can impact performance efficiency.

\textbf{Combination Stage}: 
The combination stage in GNNs involves the integration of node features with neighborhood information. This process is computationally intensive and typically comprises dense matrix multiplications, nonlinear activations, and dimensionality reduction operations~\cite{jaxed2021}. Architecturally, this stage demands high memory bandwidth and efficient data reuse mechanisms to handle large matrices. It also necessitates a balance between compute utilization and memory access, as the combination of features from large graphs can lead to memory bottlenecks.
While prior accelerators~\cite{zhang2021gamma, zhang2020sparch, li2023spada} often focus on sparse matrix multiplication tasks, they do not adequately address dense workload demands. Our NeuraChip accelerator model provides a more generalized solution, addressing the needs of both sparse graph computations and dense workloads. This approach positions NeuraChip as a versatile GNN accelerator, adept at handling both the aggregation and combination stages.

\subsection{Sparse Matrix Mult: Algorithmic Overview}
The Sparse General Matrix-Matrix Multiplication (SpGEMM) kernel execution is characterized by two main stages: the multiplication stage and the accumulation stage as visualized in \figref{fig:spgemm_stages}. The implementation variations in these stages lead to distinct SpGEMM algorithms. We describe the four approaches to execute the initial multiplication stage, as illustrated in Figure~\ref{fig:matmul_types}. These approaches vary in their memory access patterns and the level of parallelism they expose.

The inner product approach, incorporated in InnerSP~\cite{baek2021innersp} computes elements of the output matrix directly but is hindered by inefficient input reuse. Conversely, the outer product approach, utilized in OuterSPACE accelerator~\cite{pal2018outerspace} is hampered by suboptimal output locality due to the creation of numerous batches of intermediate partial product matrices~\cite{zhang2020sparch}. Our research adopts the row-wise multiplication approach (i.e., Gustavson's algorithm), selected for the extensive parallelism it provides. Notably, this approach efficiently avoids the memory bloat issue associated with handling numerous intermediate partial products~\cite{zhang2020sparch}.

The subsequent stage, known as the accumulation stage, merges the generated intermediate partial products. Various accumulation methods include heap-based~\cite{azad2016exploiting}, hash-based~\cite{nagasaka2017high}, sparse accumulator (SPA) based~\cite{gilbert1992sparse}, comparator array based~\cite{zhang2020sparch}, and Forwarding Adder Network (FAN) based~\cite{qin2020sigma,stift-jetc23}, among others (illustrated in~\figref{fig:spgemm_stages}). This stage can also be subdivided into on-chip and off-chip accumulation, based on the utilized memory hierarchy.
NeuraChip merges partial products using on-chip accumulation to reduce redundant main memory data fetches. 
For sparse matrices with irregular non-zero distributions, the on-chip accumulation stage can result in uneven workload distribution, a factor that significantly impacts the overall performance and efficiency of SpGEMM operations.


\subsection{Mapping Algorithms Design}

Mapping algorithms play a crucial role in efficiently handling computational tasks, particularly in scenarios involving sparse data structures such as those found in Graph Neural Networks (GNNs). These algorithms are tasked with assigning tasks or data elements to computational nodes or memory locations. The key requirements for effective mapping algorithms include:

\textbf{Consistency}: The algorithm must consistently map the same index to the same node. This ensures correctness in data processing.

\textbf{Low Computational Overhead}: The lookup process should be relatively fast, with minimal computational and memory overheads.
This efficiency facilitates cost-effective index matching, streamlining partial product reduction.

\textbf{Sparsity Agnostic}: Regardless of irregular sparsity patterns, the mapping algorithm should remain impartial to these variations. This ensures uniform performance across different data sets~\cite{camacho2008strong}.

Given these requirements, hash-based mapping emerges as a viable solution~\cite{shivdikar2021smash, chi2017hashing}. However, traditional hash-based methods such as Round Robin Hashing (or Ring Hashing)~\cite{takenaka2004adaptive} and Prime Number Based Modular Hashing~\cite{bhullar2016novel} have limitations~\cite{cao2000performance}. Neither is fully insensitive to sparsity patterns; a specific set of indices might consistently map to the same node, leading to potential workload imbalance.

An alternative approach is random mapping, which ideally achieves sparsity-agnostic mapping by randomly distributing indices. However, to ensure consistency, this method requires maintaining a large lookup table, which is not practical due to memory constraints.

To address these challenges, we propose a novel approach: Dynamic Reseed Hash-Based Mapping (DRHM). This method is similar to prime modular hashing, but with a significant enhancement. After processing a predetermined set of computations, we reseed the hash function. The updated seed values are then stored in a compact lookup table. This dynamic reseeding ensures that the distribution of indices does not become predictable, effectively mimicking the sparsity-agnostic property of random mapping.

Dynamic Reseed Hash-Based Mapping strikes a balance between the ideal characteristics of random mapping and the practical limitations of traditional hash-based methods. By only storing seed values rather than the entire mapping of indices, it maintains a small memory footprint. Concurrently, it offers the sparsity-agnostic mapping necessary for handling diverse and irregular data sets efficiently. This method significantly enhances the performance of computational tasks, particularly in environments where data sparsity and distribution can vary widely.

\begin{figure}[tbp]
	\centering
	\includegraphics[width=0.48\textwidth]{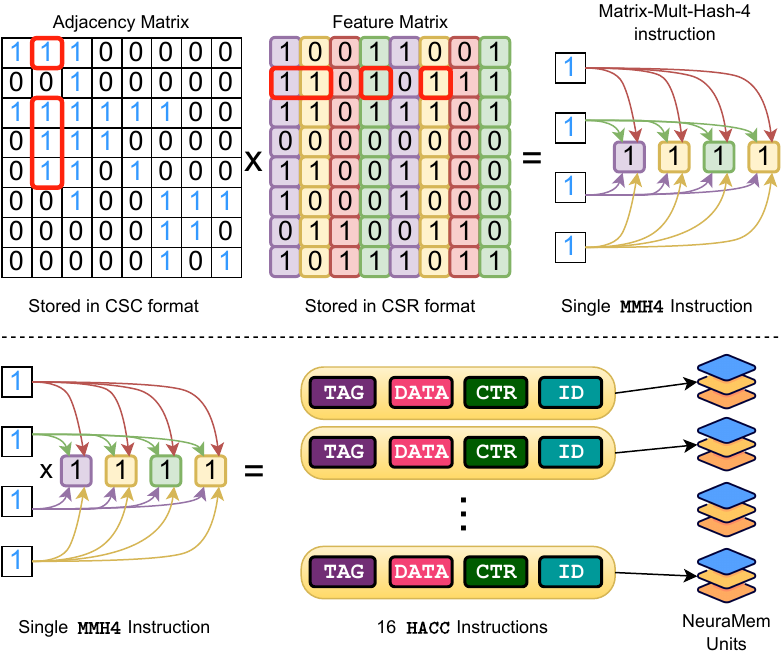}
	\caption{Implementation of tiled Gustavson's algorithm using NeuraCore for multiplication and NeuraMem for accumulation.}
	\label{fig:tiled_algorithm}
\end{figure}







%% file: sections/03_neurachip.tex
\begin{figure*}[tbp]
	\centering
	\includegraphics[width=0.96\textwidth]{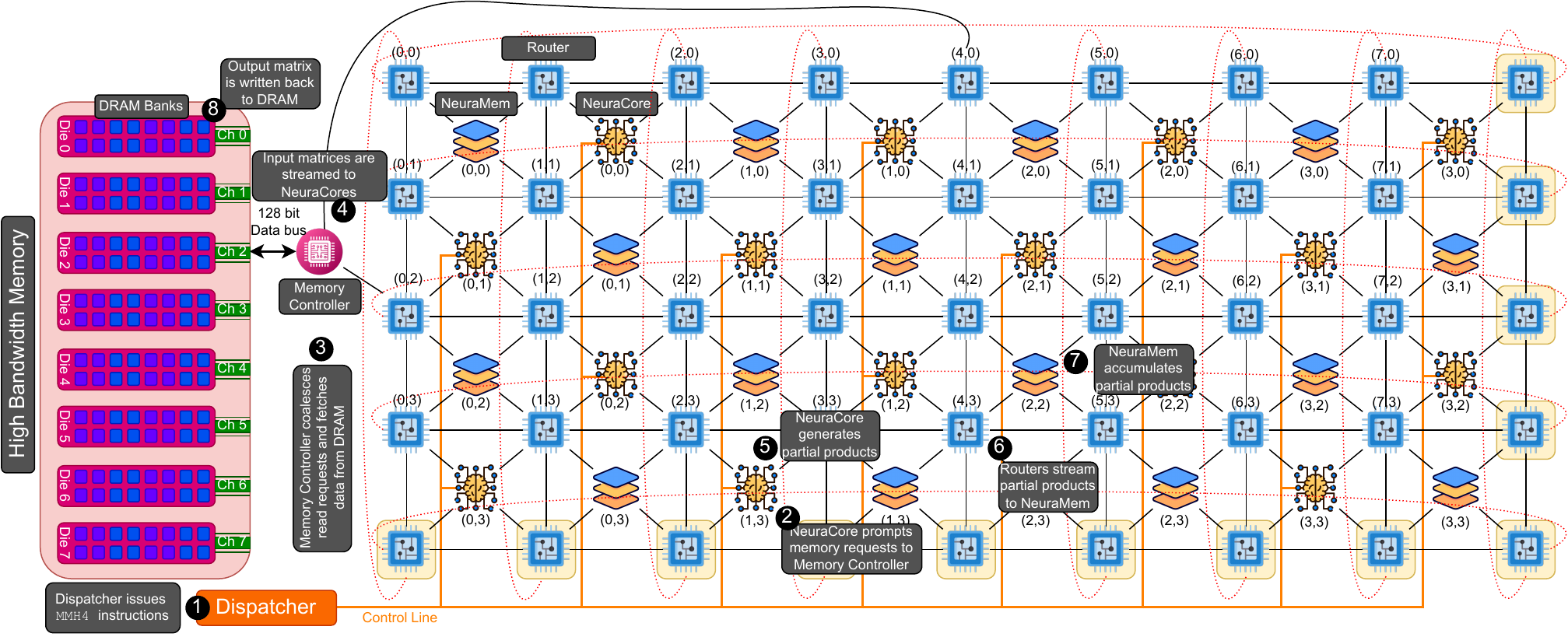}
	\vspace{-1.0em}
	\caption{NeuraChip Architecture: Tile 64 configuration with 16 NeuraCores and 16 NeuraMems per tile.}
	\label{fig:tile_16}
\vspace{-1.2em}
\end{figure*}

\newpage
\section{NeuraChip Architecture}
\label{sec:neurachip}

NeuraChip is a decoupled spatial accelerator.
Its two primary components include: i) the NeuraCore and ii) the NeuraMem. 
The NeuraCore is specifically tailored for multiplication tasks, whereas the NeuraMem focuses on accumulating data on-chip. 
They are arranged in an interleaved pattern and connected through a 2D torus network fabric, as shown in Figure~\ref{fig:tile_16}. 
To facilitate efficient communication among these components, on-chip routers have been incorporated. 
NeuraCores and NeuraMems are organized into clusters known as tiles. 
The accelerator includes a total of eight tiles, each linked to a single Double Data Rate (DDR) channel. 
Each tile features a memory controller that interfaces with DRAM banks.

Buffers play a critical role in the functionality of the four major components of our accelerator. 
Both the NeuraCore and the NeuraMem are equipped with instruction buffers. 
Additionally, the on-chip routers incorporate packet buffers, and the memory controllers are fitted with buffers for managing both reading and writing operations.

The incorporation of these on-chip buffers enhances the accelerator's flexibility, allowing it to adapt to diverse sparsity patterns. 
In scenarios where irregular graph structures could lead to network congestion, these on-chip buffers prove beneficial. 
They ensure that the components consistently have instructions to execute, thus avoiding potential delays or bottlenecks in processing.

\subsection{Tiled Gustavson's Multiplication Algorithm}

GNNs typically employ two primary layers (phases) in their architecture: the neighborhood aggregation phase, which gathers information from a node's neighbors in the graph, and the combination phase, where a node's representation is updated by integrating its own features with those aggregated from its neighbors. 
This discussion focuses on the aggregation phase, which predominantly involves sparse matrix multiplications~\cite{hamilton2020graph}.

In this paper, we implement a modified version of Gustavson's matrix multiplication algorithm~\cite{gust1978matmul}. 
Gustavson's algorithm operates on a row-stationary approach, processing the output matrix one row at a time. 
Specifically, it traverses the adjacency matrix row by row, performing a linear combination of these rows as illustrated in~\figref{fig:tiled_algorithm}.

Gustavson's approach multiplies each element in a row of the adjacency matrix with all elements in the corresponding row in the feature matrix that has the same row index as the element's column index. 
Our adaptation enhances Gustavson's method by simultaneously processing multiple rows. 
We execute the multiplication of four rows at a time, aligning four elements from a column of the adjacency matrix with four elements from a row of the feature matrix.
This is achieved using a specialized instruction, denoted as the \(\texttt{MMH4}\) instruction.

Our technique represents a fusion of Gustavson's algorithm and the outer-product method.
Unlike the outer-product approach which finalizes the multiplication of an entire column with a row before moving to the next, our strategy concurrently processes four rows by employing the Gustavson method. 
The selection of the number `four' for simultaneous row processing results from design space exploration specific to the NeuraChip accelerator. 

To implement this modified Gustavson's approach, the adjacency matrix is stored in a compressed sparse column (CSC) format, and the feature matrix is stored in a compressed sparse row (CSR) format.
However, this approach presents two primary challenges:

\textbf{Unavoidable Index Matching}: Employing Gustavson's algorithm and compressed matrix storage formats such as CSR and CSC inherently leads to the necessity of index matching~\cite{srivastava2020matraptor, peng2023maxk}. 
We address the index-matching overhead with a constant lookup hash function, facilitating the on-chip accumulation of partial products with a constant lookup time. 
The low overhead provided by our hash function is further optimized by adding a dedicated hash engine, as described in \secref{ssec:neuramem}.

\textbf{Memory Bloat Issue}: The tiled Gustavson method can result in memory bloat, characterized by the generation of a large number of partial products. 
To tackle this issue, we have implemented a rolling eviction mechanism. 
This system accumulates partial products as they are generated and promptly evicts them once the reduction is complete, with further details provided in Section~\ref{ssec:neuramem}.

\subsection{On-chip Dataflow}

To illustrate the data flow within NeuraChip, we walk through an example of an SpGEMM kernel executed on the NeuraChip accelerator (see Figure~\ref{fig:tile_16}).
Step \circled{1} The process begins with the \textit{Dispatcher} issuing \textbf{matrix\_mult\_hash\_4} (\texttt{MMH4}) instructions to every \textit{NeuraCore}.
Step \circled{2} The \textit{NeuraCores} trigger memory read requests that are routed to the memory controller.
Step \circled{3} The \textit{Memory Controller} coalesces requests for contiguous memory locations into a singular transaction and reorganizes memory transactions to enhance spatial locality.
Step \circled{4} Input matrix data, fetched from DRAM, is streamed onto respective \textit{NeuraCore} components.
Step \circled{5} The \textit{NeuraCores} compute the partial products, along with their corresponding rolling counters (further details in Section~\ref{ssec:neuracore}), subsequently generating the \textbf{hash\_accumulate} (\texttt{HACC}) instructions.
Step \circled{6} \texttt{HACC} instructions are streamed over on-chip routers into NeuraMem components, based on a hash-based mapping.
Step \circled{7} The \textit{NeuraMem} component employs another hash function to hash and accumulate these partial products onto their on-chip memory. 
Consecutive hashes of partial products with the same TAG are merged within NeuraMem, with each hash insertion decrementing the counter by $1$.
Step \circled{8} When the counter reaches zero; this triggers the eviction of the hashline, and the resultant data is written back to the High Bandwidth Memory (HBM).

\subsection{NeuraCore}
\label{ssec:neuracore}


The NeuraCore is the primary compute engine in our accelerator. It computes the multiplication operation and generates corresponding partial products. It is a simple in-order core with support for matrix instructions.
NeuraCore supports a special matrix instruction called \texttt{matrix\_mult\_hash\_4} or simply \texttt{MMH4}.

\begin{algorithm}[b]
\caption{\texttt{MMH4} instruction execution}
\label{algo:mmh4}
\begin{algorithmic}[1]
\FOR{$i = 0$ \TO $3$}
    \FOR{$j = 0$ \TO $3$}
        \STATE $TAG \gets \text{Mem}[(Base_{addr} + B_{col\_ind\_addr} + j)]$
        \STATE $DATA \gets \text{Mem}[(Base_{addr} + A_{data\_addr} + i)]$
        \STATE \hspace{\algorithmicindent} $\times \text{Mem}[(Base_{addr} + B_{data\_addr} + j)]$
        \STATE $CTR \gets \text{Mem}[(Base_{addr} + roll\_counter + i*4 + j)]$%
        \STATE $\text{Dispatch} \ \texttt{HACC}(TAG, DATA, COUNTER)$
    \ENDFOR
\ENDFOR
\end{algorithmic}
\end{algorithm}

\begin{figure}[bp]
	\centering
        \vspace{-1.3em}
	\includegraphics[width=0.48\textwidth]{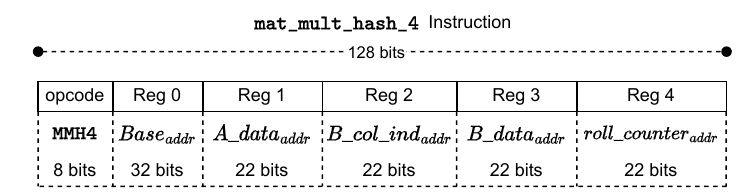}
	\vspace{-0.6em}
	\caption{\texttt{MMH4} instruction bit layout.}
	\label{fig:mmh4}
\vspace{-1.7em}
\end{figure}

Algorithm~\ref{algo:mmh4} presents the \texttt{MMH4} instruction execution pseudocode where
$opcode$ represents the operation code, which specifies the MMH4 instruction to be executed by NeuraCore.
$Base_{addr}$ denotes the base address used to offset the address of all other addresses involved in this instruction.
$A\_data_{addr}$ refers to the memory address where the data of matrix A is located (matrix A is stored in CSC storage format).
$B\_col\_ind_{addr}$ points to the memory address containing the column indices of matrix B (matrix B is stored in CSR storage format).
$B\_data_{addr}$ indicates the memory address where data from matrix B is stored.
$roll\_counter_{addr}$ denotes the memory address where the rolling eviction counter is located.
The instruction layout for \texttt{MMH4} is presented in Figure~\ref{fig:mmh4}. Each \texttt{MMH4} instruction has the capability to dispatch up to $16$ \texttt{HACC} instructions (further elaborated in the NeuraMem section).



\begin{figure}[tbp]
	\centering
	\includegraphics[width=0.47\textwidth]{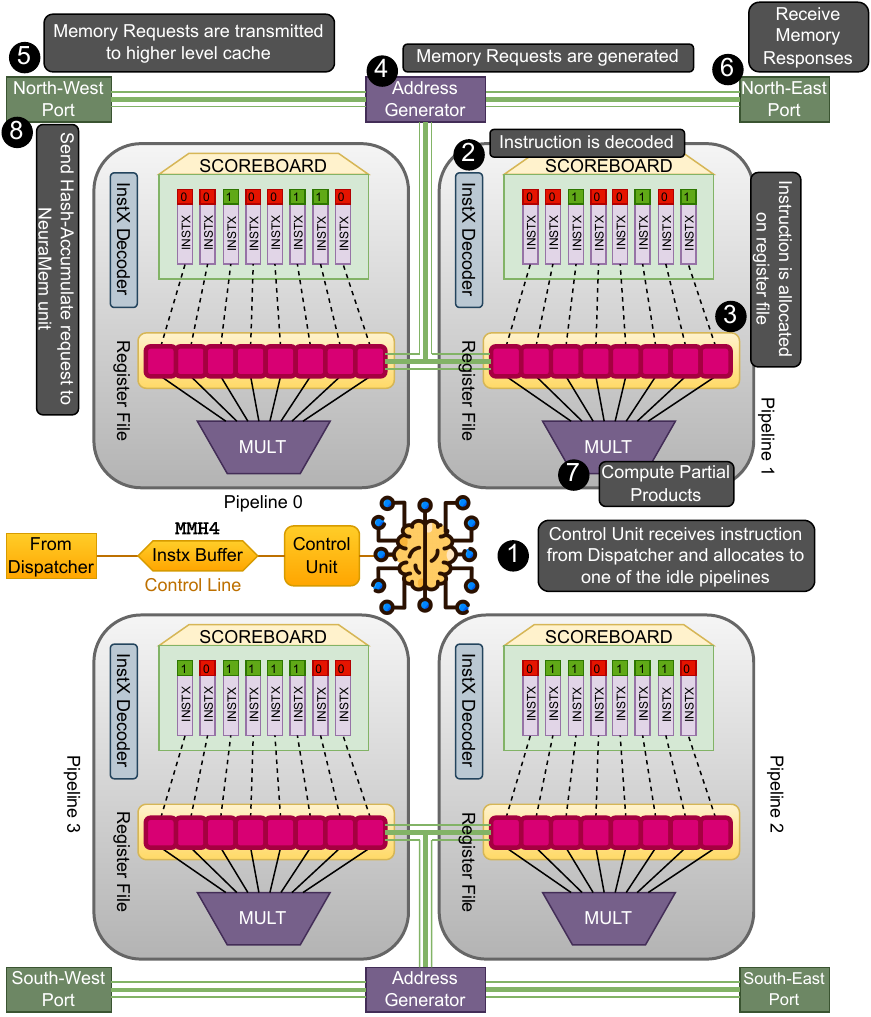}
	\caption{NeuraCore's quad-pipeline layout.}
	\label{fig:neuracore}
\vspace{-1.7em}
\end{figure}

The operational sequence within NeuraCore is shown in Figure~\ref{fig:neuracore}, and can be broken down into the following steps:
Step \circled{1}: The operation starts with the dispatcher transmitting a \texttt{MMH4} instruction to NeuraCore, allocating the instruction to one of the available pipelines. Pipelines are allocated using a round-robin scheme.
Step \circled{2}: The \texttt{MMH4} instruction is decoded by the on-chip decoder.
Step \circled{3}: Following decoding, NeuraCore maps instruction variables to the register file, utilizing dynamic register allocation.
Step \circled{4}: Post register allocation, the NeuraCore's internal address generator constructs memory requests to fetch elements from the input matrices.
Step \circled{5}: An adaptive routing algorithm~\cite{ascia2008implementation} selects the best port to dispatch the memory request, which is then forwarded to a higher-level cache.
Step \circled{6}: Upon completing the memory request, a response is received at one of the NeuraCore's four ports. This response is then routed toward its respective pipeline.
Step \circled{7}: As soon as all memory responses corresponding to a particular instruction are received, the instruction is deemed ready for execution by the scoreboard. Subsequently, the multiplication pipeline calculates the partial product and generates up to $16$ \texttt{HACC} instructions.
Step \circled{8}: Lastly, the \texttt{HACC} instructions are relayed to NeuraMem units using the most suitable port, as determined by the on-chip hash-based mapping function.


\subsection{NeuraMem}
\label{ssec:neuramem}

\begin{figure}[tbp]
	\centering
	\includegraphics[width=0.48\textwidth]{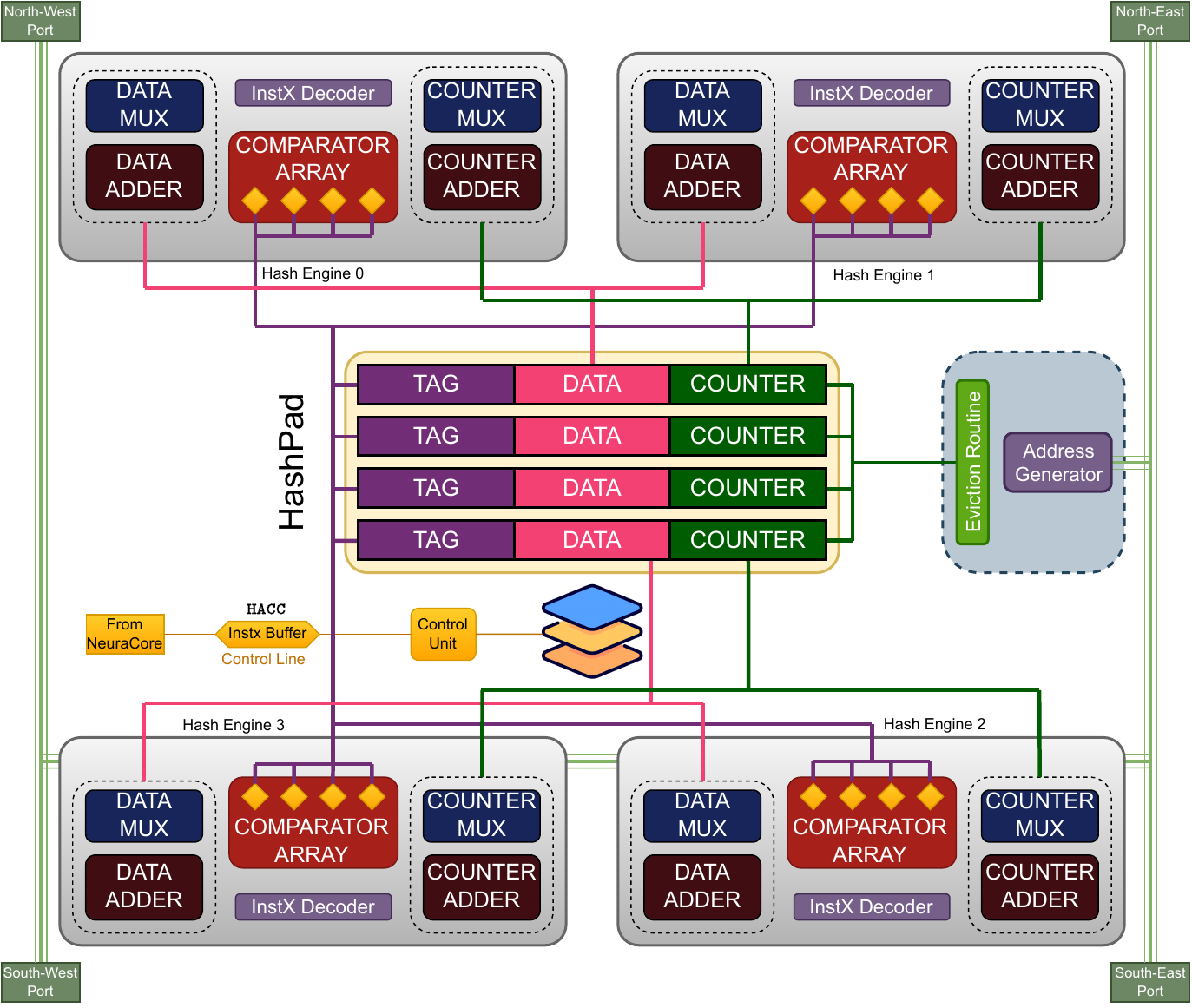}
	\vspace{-2.2em}
	\caption{NeuraMem's quad-hash-engine block diagram.}
	\label{fig:neuramem}
\vspace{-1.7em}
\end{figure}

\begin{figure}[bp]
	\centering
 \vspace{-1.7em}
	\includegraphics[width=0.48\textwidth]{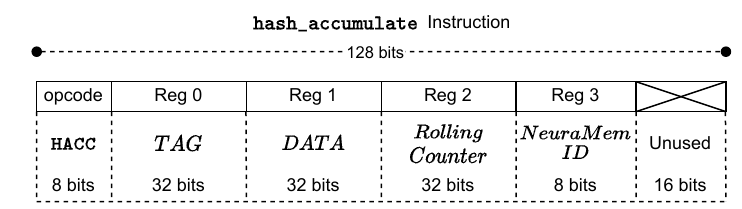}
        \vspace{-0.6em}
	\caption{\texttt{HACC} instruction bit layout.}
	\label{fig:hacc}
\end{figure}

NeuraMem is a crucial component of the NeuraChip accelerator. While NeuraCore units generate partial products, NeuraMem units handle the on-chip accumulation of these partial products. The central component of NeuraMem units is the Hash-Engine. The layout of various components within NeuraMem is as shown in Figure~\ref{fig:neuramem}.

\textbf{HashPad}: The Hash-Engine operates on what we refer to as ``hash-lines''~\figref{fig:neuramem}. A hash-line comprises a single TAG, DATA, and COUNTER entry. The collective TAG array, DATA array, and COUNTER array, essentially the whole set of hash-lines, form what is known as the HashPad, as shown in Figure~\ref{fig:neuramem}.

\textbf{\texttt{HACC} instruction}: NeuraMem supports a special instruction for partial product accumulation called \texttt{hash\_accumulate}, or simply \texttt{HACC} instruction. The bit layout of \texttt{HACC} instruction is illustrated in \figref{fig:hacc}. \algoref{algo:hacc} presents a pseudocode of the \hacc  instruction, providing clearer insight into its functionality.


\begin{figure}[tbp]
	\centering
	\includegraphics[width=0.40\textwidth]{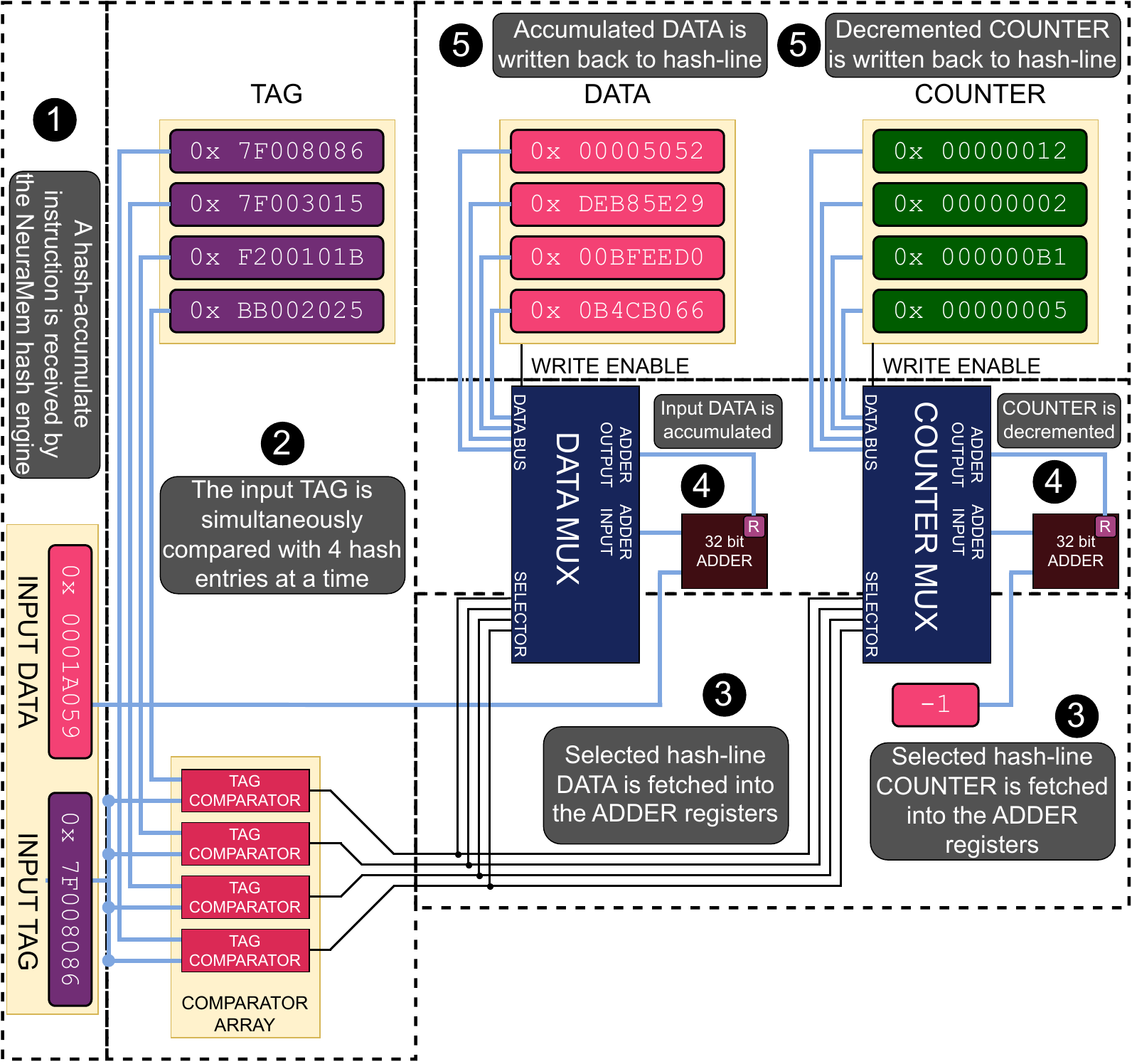}
	\vspace{-1.0em}
	\caption{NeuraMem Hash-Engine accumulates partial products using the \hacc instruction.}
	\label{fig:hash_engine}
\vspace{-1.7em}
\end{figure}

\textbf{Hash-Engine workflow}: \figref{fig:hash_engine} shows a typical sequence of events during the execution of a \texttt{HACC} instruction by the Hash-Engine (illustrated using pseudocode in \algoref{algo:hacc}). The process starts in step \circled{1}, where the Hash-Engine receives a \hacc instruction from the NeuraCore units. This instruction's \mtag is simultaneously compared with all the TAGs currently present on the HashPad (step \circled{2}). The multiplexers select the hash-line with the matching TAG in step \circled{3}. The corresponding hash-line's DATA gets accumulated with the \hacc instruction's data. Simultaneously, the counter for that hash-line is decremented by one (step \circled{4}). The accumulated data and the updated counter are then written back to the HashPad in step \circled{5}.
If the TAG from the instruction does not match any of the TAGs in the HashPad in step \circled{2}, the Hash-Engine creates a new entry for the hash instruction and stores its content in a new hash-line.

\begin{algorithm}[b]
\caption{\texttt{HACC} Instruction Execution}
\label{algo:hacc}
\begin{algorithmic}[1]
\STATE $index \gets \text{Hash}(TAG)$
\IF{$tag\_array[index] == \text{EMPTY}$} 
    \STATE $data\_array[index] \gets DATA$
    \STATE $counter\_array[index] \gets COUNTER$
\ELSIF{$tag\_array[index] == TAG$}
    \STATE $data\_array[index] \mathrel{+}= DATA$
    \STATE $counter\_array[index] \mathrel{-}= 1$
    \IF{$counter\_array[index] == 0$}
        \STATE $\text{Hash Line Eviction Routine}$
    \ENDIF
\ELSE
    \STATE $\text{Hash Collision Routine}$
\ENDIF
\end{algorithmic}
\end{algorithm}


\textbf{Rolling Evictions}: The Hash-Engine monitors the completion of partial product accumulation (via the COUNTER, as seen in \figref{fig:neuramem}. Once the COUNTER reaches zero, indicating that all partial products for a particular TAG have been accumulated, the Hash-Engine automatically evicts the corresponding hash-line, and the accumulated result is written back to the main memory (HBM). This ensures that the hashed partial product spends the minimal possible number of cycles in the HashPad, addressing the memory bloat issue.


\subsection{Dynamically Reseeding Hash-based Mapping}
\label{ssec:drhm}
%
%

The performance benefits provided by the NeuraChip accelerator are primarily due to our sparsity-agnostic mapping algorithm, named Dynamically Reseeding Hash-based Mapping (DRHM). DRHM is designed to eliminate computational patterns, promoting an even distribution of workload across all computational resources. Traditional hash-based mappings often lead to concentrated areas of high activity, known as hot spots, especially when the hash function is optimized for a specific sparsity pattern but encounters a different one. An ideal solution would involve uniformly distributing computational tasks across resources. One such method is random mapping, where tasks are allocated to random resources. However, maintaining consistency in random mapping requires extensive record-keeping (a large lookup table), which is impractical.

We introduce a hybrid approach, Dynamically Reseeding Hash-based Mapping (DRHM), which combines the advantages of consistent lookup times in hashing, a distribution akin to random mapping, and minimal overhead similar to small lookup tables. This method significantly reduces the occurrence of hot spots in the allocation of computational resources.




DRHM utilizes a flexible mapping that adjusts based on a `seed' parameter, denoted as $\gamma$. This parameter is specifically designed to alter the mapping, and consequently, the hash function dynamically. After each row of the input sparse matrix is computed, $\gamma$ is initialized with a random number.
DRHM offers two implementation approaches: one using the $k$ upper bits of the \mtag, and the other utilizing the $k$ lower bits of the \mtag. The lower-bit and upper-bit hashing equations that accommodate $\gamma$ seed are presented in Equations~\ref{eq:DRHM_lower} and \ref{eq:DRHM_higher}.


\begin{equation}
\label{eq:DRHM_lower}
H_{l}(\mathrm{TAG_{32}}, \gamma) = ((\mathrm{TAG_{32}} \ll k) \gg k) \cdot \gamma \mod N
\end{equation}
\begin{equation}
\label{eq:DRHM_higher}
H_{h}(\mathrm{TAG_{32}}, \gamma) = ((\mathrm{TAG_{32}} \gg k) \ll k) \cdot \gamma \mod N
\end{equation}

where \mtag represents the unique identifier for each row of the input graph. The term $\gamma$ acts as a `seed' to introduce randomness in the mapping. $N$ signifies the total number of available output hash spaces. The operations ``$\ll k$'' and ``$\gg k$'' refer to bitwise left and right shifts by $k$ positions, respectively. The modulus operation $\mod$ ensures that the result of the hash function falls within the predefined range of the hash table.
These equations assume that the bit-shift operations conform to standard behavior where bits shifted beyond the boundary of the number's bit-width are discarded.

\begin{figure}[tp]
	\centering
	\includegraphics[width=0.38\textwidth]{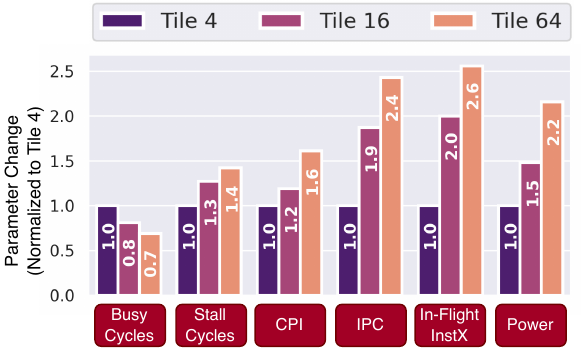}
	\vspace{-1.0em}
	\caption{Architectural impact of GCN model varying tile configuration on Cora dataset.}
	\label{fig:tile_analysis}
\vspace{-1.9em}
\end{figure}

In our experiments, we assessed both upper $k$-bit address hashing and lower $k$-bit address hashing. We found that the lower $k$-bit address hashing method had a lower incidence of hash collisions, due to the higher variability in the lower bits of the address. Consequently, in all the work presented here, we employ the lower $k$-bit address hashing technique (Equation~\ref{eq:DRHM_lower}).
Our compute mapping efficiancy using DRHM approach is evaluated in \secref{sec:dse}.

%% file: sections/04_design_space.tex
\section{Design Space Exploration of NeuraChip}
\label{sec:dse}

The flexibility of our NeuraSim simulator enables us to evaluate multiple NeuraChip configurations. We have two primary design goals: i) optimizing resource utilization across the accelerator to enhance speedup and ii) striking a balance between performance, chip area, and power consumption to make sure the advantages outweigh the costs~\cite{srinivasan2016dynamic, banerjee2004power}.

\input{tables/comp_config}

\textbf{Tile Size Variation}: We define a tile as a modular unit that can be configured with varying amounts of computational and memory resources. Each tile is connected to a dedicated HBM memory channel, with NeuraChip hosting a total of eight tiles to match the eight available memory channels.
We introduce three distinct configurations of NeuraChip, named Tile-4, Tile-16, and Tile-64, derived from experimenting with various workloads. The detailed configurations of NeuraCore and NeuraMem components are provided in \tabref{tab:comp_config}, while the overall accelerator configurations for these tile sizes are listed in \tabref{tab:chip_configuration}. We focus on six key parameters to assess the architectural impact of these configurations, as shown in \figref{fig:tile_analysis}.
Key observations include:

\begin{itemize}
    
    \item \textbf{Register File Size}: Expanding the register file size allows more \mmhf instructions to be in-flight and increases the number of read memory instructions that can be issued to HBM. Beyond $8$ registers per pipeline ($1024$ bits per pipeline), we noticed that the DRAM channels are unable to keep up with the high memory demands.
    This bottleneck is evident in the rise in the cycles per instruction (CPI) and the number of stall cycles~\cite{shivdikar2023gme, accelerating2023livesay}, as shown in \figref{fig:tile_analysis}.

    \item \textbf{HashPad Size}: Choosing between smaller HashPads with a larger number of NeuraMems versus larger HashPads with fewer NeuraMems, the former proves advantageous for handling extremely sparse matrices. This configuration benefits from high accumulation throughput as the number of accumulators increases with the number of NeuraMems. This can be seen in the larger number of in-flight HBM memory instructions in \figref{fig:tile_analysis}.

    \item \textbf{Component Counts}: With 32, 128, and 512 NeuraCores and NeuraMems in Tile-4, Tile-16, and Tile-64, respectively, while more components enhance peak compute throughput, the configuration is bound by a peak DRAM bandwidth of $128\ GB/s$. Additionally, workloads do not require a 12 MB on-chip memory  HashPad (of tile-64 configuration).

\end{itemize}

\input{tables/chip_configuration}

\begin{figure}[bp]
	\centering
	\includegraphics[width=0.4\textwidth]{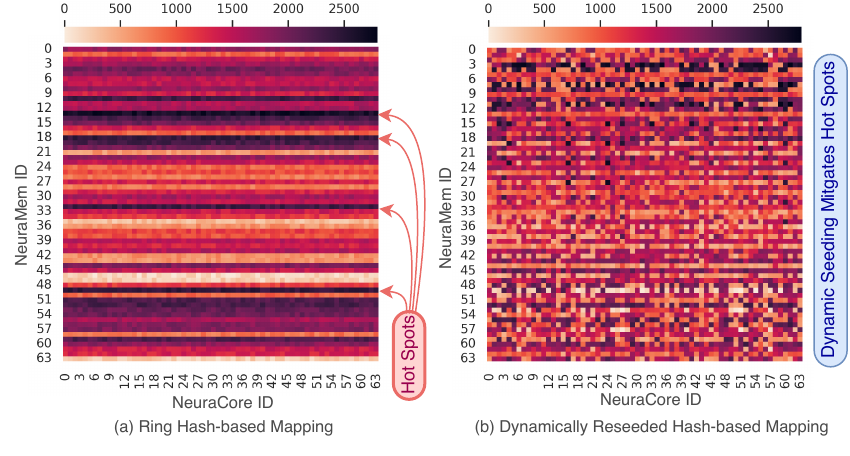}
	\vspace{-1.2em}
	\caption{Compute mapping heat map, where the X-axis represents multiplications mapped to NeuraCores and Y-axis represents accumulations mapped to NeuraMem.}
	\label{fig:hop_dynamic}
\vspace{-1.5em}
\end{figure}

\textbf{Hash-based Mapping Algorithm Variations}: We tested four hash-based mapping schemes. The first, a ring-based mapping (see \figref{fig:hop_dynamic}), follows round-robin resource allocation, though encounters hot spots in workload distribution. The second, a modular hash-based mapping, uses prime numbers for workload mapping, proposed in previous studies~\cite{zhang2018fast, gou2018single, ng2015two, song2009design}. DRHM, shown in \figref{fig:hop_dynamic}, addresses hot spots in modular and ring-based mappings by reseeding the hash function after each row of computations. Lastly, we evaluate a random mapping that maintains a lookup table for each entry. All four techniques are compared in \figref{fig:dynamic_hashing} for varying sparsity patterns.

\textbf{Variations in \texttt{MMH} and \texttt{HACC} Instructions}: 
NeuraChip introduces \texttt{MMH} and \texttt{HACC} instructions (bit layout of these instructions is illustrated in~\figref{fig:mmh4} and~\figref{fig:hacc}), supporting its decoupled architecture~\cite{shivdikar2018speeding}. We analyze the cycle count of various \texttt{MMH} instruction tile sizes, presented in a CPI histogram in \figref{fig:mmh_cpi}. \texttt{MMH4} emerges as the top choice, balancing temporal locality benefits and cycle count.

We compare the \texttt{HACC} instruction's efficiency using two eviction schemes: barrier-based eviction (\texttt{HACC-BE}) and our rolling eviction approach (\texttt{HACC-RE}). The latter's superiority in reducing average cycle completion is seen in \figref{fig:hacc_cpi}.

%% file: tables/comp_config.tex
\begin{table}
\centering
\caption{Individual Component Configuration}
\vspace*{-2mm}
\label{tab:comp_config}
\begin{tabular}{p{0.6in} p{0.9in} x{0.4in} x{0.4in} x{0.4in}}
\arrayrulecolor{black}\toprule
    \textbf{Component}&  \textbf{Elements} & \textbf{Tile-4} & \textbf{Tile-16} & \textbf{Tile-64} \\ [0.5ex] 
\arrayrulecolor{black}\toprule
    \multirow{5}{*}{\textbf{NeuraCore}} & Pipeline Registers & $4$ & $8$ & $16$ \\
                                        & Pipelines & $2$ & $4$ & $8$ \\
                                        & Multipliers & $2$ & $4$ & $8$ \\
                                        & Addr. Generators & $1$ & $2$ & $2$ \\
                                        & Ports & $4$ & $4$ & $4$ \\                                        
    \arrayrulecolor{black!20}\midrule
    \multirow{5}{*}{\textbf{NeuraMem}}  & Comparators & $1$ & $4$ & $8$ \\
                                        & Hash-Engines & $2$ & $4$ & $8$ \\    
                                        & Hashlines & $4096$ & $2048$ & $2048$ \\
                                        & Accumulators & $128$ & $256$ & $512$ \\
                                        & Ports & $4$ & $4$ & $4$ \\

     \arrayrulecolor{black}\bottomrule

\end{tabular}




\vspace{-5mm}
\end{table}

%% file: tables/chip_configuration.tex
\begin{table}
\centering
\caption{NeuraChip Configuration}
\label{tab:chip_configuration}
\begin{tabular}{p{1.4in} | x{0.4in} x{0.5in} x{0.5in}}
\arrayrulecolor{black}\toprule
   \textbf{Parameter} & \textbf{Tile-4}$^*$ & \textbf{Tile-16}$^*$ & \textbf{Tile-64}$^*$ \\ [0.5ex] 
\arrayrulecolor{black}\toprule
\arrayrulecolor{black}\toprule

     Tile Count & $8$ & $8$ & $8$ \\
    \arrayrulecolor{black!20}\midrule
    NeuraCores per tile & $1$ & $4$ & $16$ \\
    \arrayrulecolor{black!20}\midrule
    Total NeuraCores & $8$ & $32$ & $128$ \\
    \arrayrulecolor{black!20}\midrule
    NeuraMems per tile & $1$ & $4$ & $16$ \\
    \arrayrulecolor{black!20}\midrule
    Total NeuraMems & $8$ & $32$ & $128$ \\
    \arrayrulecolor{black!20}\midrule
    Memory Controller Count & $8$ & $8$ & $8$ \\
    \arrayrulecolor{black!20}\midrule
    Routers per tile & $4$ & $8$ & $32$ \\
    \arrayrulecolor{black!20}\midrule
    Total Routers & $32$ & $64$ & $256$ \\
    \arrayrulecolor{black!20}\midrule
    Total Pipelines & $32$ & $128$ & $512$ \\
    \arrayrulecolor{black!20}\midrule
    Pipeline Register File (bits) & $512$ & $1024$ & $2048$ \\
    \arrayrulecolor{black!20}\midrule
    Total Hash-Engines & $16$ & $128$ & $1024$ \\
    \arrayrulecolor{black!20}\midrule
    Hash-Engine comparators & $2$ & $4$ & $8$ \\
    \arrayrulecolor{black!20}\midrule
    Total \mtag comparators & $32$ & $512$ & $8192$ \\
    \arrayrulecolor{black!20}\midrule
    Total HashPad Size (MB) & $0.75$ & $3$ & $12$ \\
    \arrayrulecolor{black!20}\midrule
    Max frequency ($GHz$) & $1$ & $1$ & $1$ \\

     \arrayrulecolor{black}\bottomrule
\end{tabular}



\vspace{-0.8em}
\end{table}


%% file: sections/05_evaluation.tex
\section{Evaluation}

\subsection{Experimental Setup}


To evaluate the benefits of NeuraChip, we perform benchmarking across two distinct categories of workloads.
The first category involves examining NeuraChip's efficiency in handling sparse matrix multiplication tasks.
This evaluation uses a standard array of sparse matrices obtained from the Stanford SNAP sparse matrix collection~\cite{snapnets}.
Our evaluation includes a comparison with some of the latest state-of-the-art sparse matrix accelerators~\cite{zhang2020sparch, zhang2021gamma} and off-the-shelf mainstream hardware platforms. NeuraChip is benchmarked against the Intel MKL library~\cite{wang2014intel} with an Intel Xeon E5-2630 CPU. We also compare against cuSPARSE~\cite{naumov2010cusparse} and CUSP~\cite{Cusp} NVIDIA libraries, as run on a Hopper architecture H100 GPU, and we also consider for comparison an AMD's MI100 GPU using the hipSPARSE library with a rocSPARSE backend~\cite{ROCmSoftwarePlatform}.
For accelerator comparisons, we compare NeuraChip against OuterSPACE~\cite{pal2018outerspace} SpArch~\cite{zhang2020sparch}, and Gamma~\cite{zhang2021gamma}.
Additionally, as to the second category of workloads, our evaluation targets a Graph Convolutional Network (GCN)~\cite{kipf2016semi} layer using various datasets, allowing us to compare NeuraChip against existing Graph Neural Network (GNN) accelerators EnGN~\cite{liang2020engn}, GROW~\cite{hwang2023grow}, HyGCN~\cite{yan2020hygcn}, and FlowGNN~\cite{sarkar2023flowgnn}.

\begin{figure*}[tbp]
	\centering
	\includegraphics[width=0.65\textwidth]{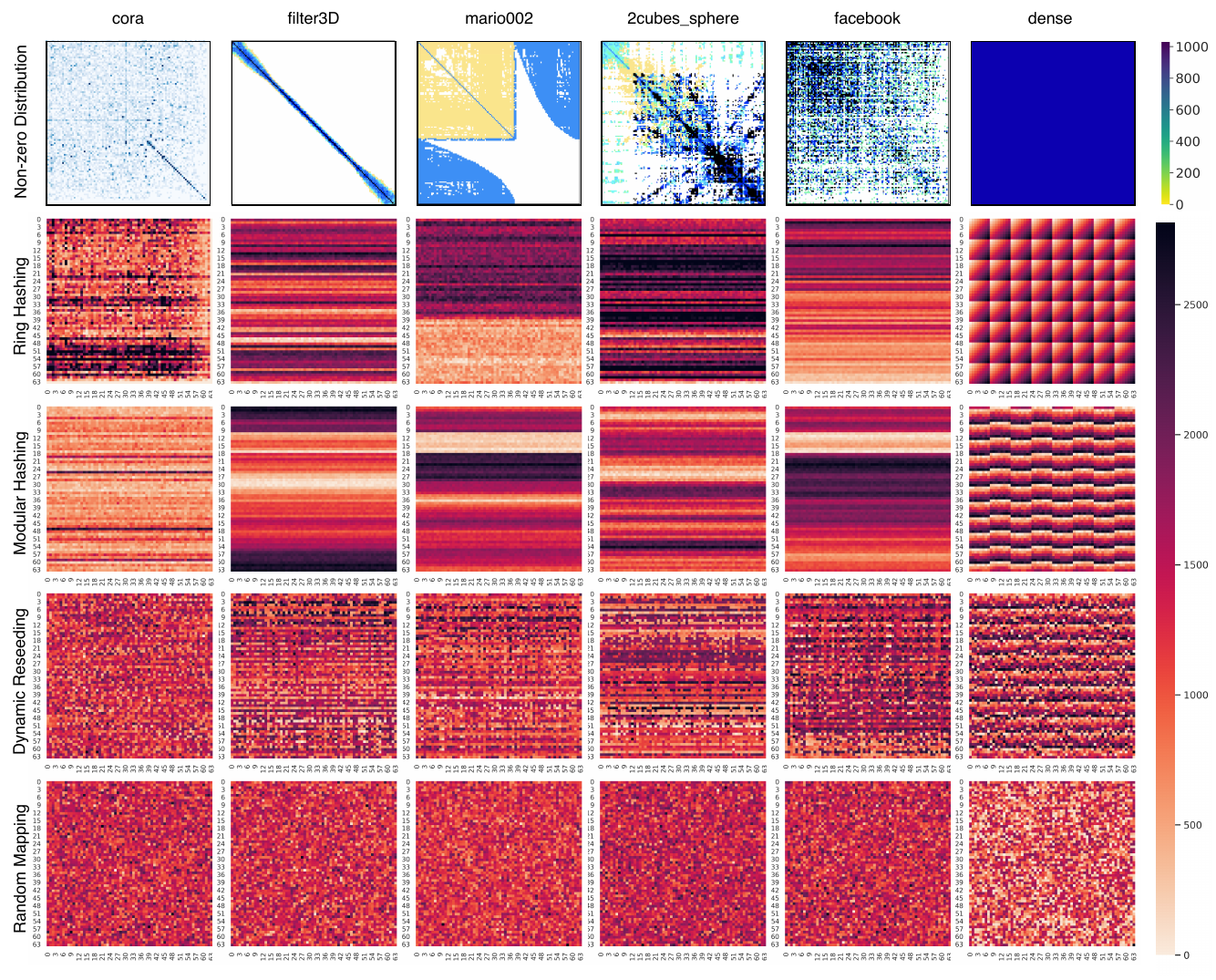}
	\vspace{-1.2em}
	\caption{Computation mapping heat maps for four distinct hash-based mapping methods, evaluated across five sparse matrices and one dense matrix multiplication. The dynamic reseeding mapping technique is insensitive to sparsity patterns and effectively addresses hot spots in dense matrix computations.}
	\label{fig:dynamic_hashing}
\vspace{-1.7em}
\end{figure*}

\begin{figure}[bp]
	\centering
 \vspace{-1.7em}
	\includegraphics[width=0.48\textwidth]{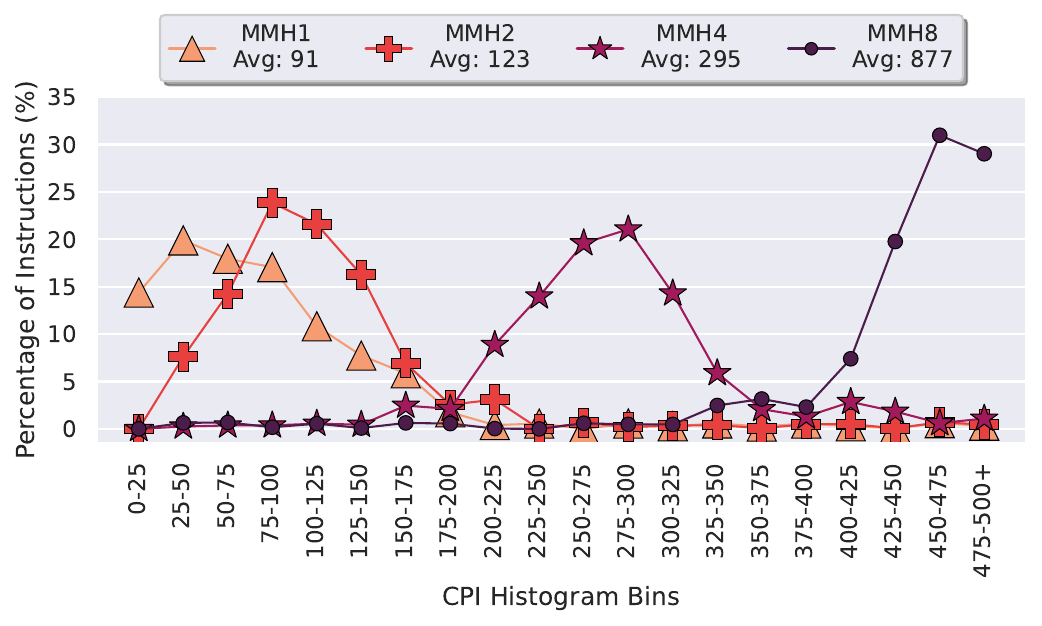}
	\vspace{-1.2em}
	\caption{Cycles Per Instruction (CPI) histogram plot for four \texttt{MMH} instructions with varying tile sizes.}
	\label{fig:mmh_cpi}
\vspace{-1.7em}
\end{figure}

\subsection{Simulator Framework}

In this study, we present NeuraSim, a cycle-accurate, multi-threaded, modular simulation engine inspired by the Structural Simulation Toolkit (SST)~\cite{rodrigues2011structural}. NeuraSim's modular framework allows for flexible integration of new architectural features, without the need for an entire overhaul of the simulation engine. Developed using POSIX threads (\texttt{pthreads}), NeuraSim facilitates parallel simulation. Its dispatcher unit recognizes independent tasks and concurrently executes them on different threads. Additionally, NeuraSim employs MongoDB for backend data storage. NeuraSim also incorporates HBM2 memory simulation, integrating with DRAMsim3~\cite{li2020dramsim3}, a cycle-accurate and validated DRAM simulator.

Regarding simulation efficiency, NeuraSim achieves $112$ Kilocycles per second (KCPS), $48$ KCPS, and $11$ KCPS on average for the Tile-4, Tile-16, and Tile-64 configurations, respectively. NeuraSim is open-source and faithfully simulates the extended NeuraChip ISA. The NeuraSim source code is accessible on our GitHub repository.

\subsection{Comparative Analysis with Sparse Matrix Accelerators}

\begin{figure*}[tbp]
	\centering
	\includegraphics[width=0.98\textwidth]{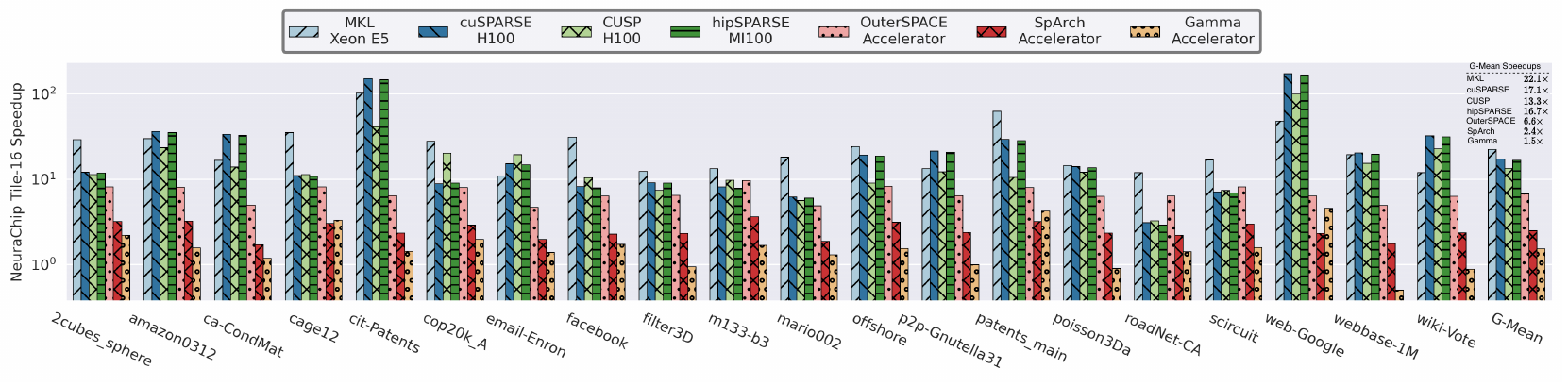}
	\vspace{-1.2em}
	\caption{Speedup comparison of NeuraChip Tile-16 for SpGEMM against CPUs, GPUs, and SpGEMM accelerators.}
	\label{fig:speedup}
\vspace{-1.0em}
\end{figure*}

In \figref{fig:speedup}, the performance of the NeuraChip in sparse matrix multiplication tasks is compared against various off-the-shelf high-end CPU and GPU platforms, as well as against state-of-the-art SpGEMM accelerators.

As we can see, NeuraChip outperforms the CPU and GPU computing platforms in all cases. The average performance improvements are $22.2\times$ over the CPU, a $17.1\times$ and $13.3\times$ average speedup over the NVIDIA Hopper GPU using the cuSPARSE and CUSP libraries, respectively, and $16.7\times$ average speedup over the AMD's MI100 GPU using hipSPARSE.

Further, the performance of NeuraChip is evaluated against two outer-product-based sparse matrix accelerators: OuterSPACE~\cite{pal2018outerspace} and SpArch~\cite{zhang2020sparch}. While OuterSPACE leverages input data reuse, it encounters excessive generation of partial products (the memory bloat issue), leading to degraded performance. SpArch addresses this with on-chip merger trees; however, these trees require large comparator arrays, occupying about $60\%$ of the chip area. NeuraChip counters the memory bloat through an on-chip cache organization with rolling counters, effectively managing the eviction of accumulating partial products and alleviating the bloat issue. In comparison, NeuraChip surpasses OuterSPACE and SpArch by factors of $6.6\times$ and $2.4\times$, respectively.

\begin{figure}[bp]
	\centering
 \vspace{-1.7em}
	\includegraphics[width=0.48\textwidth]{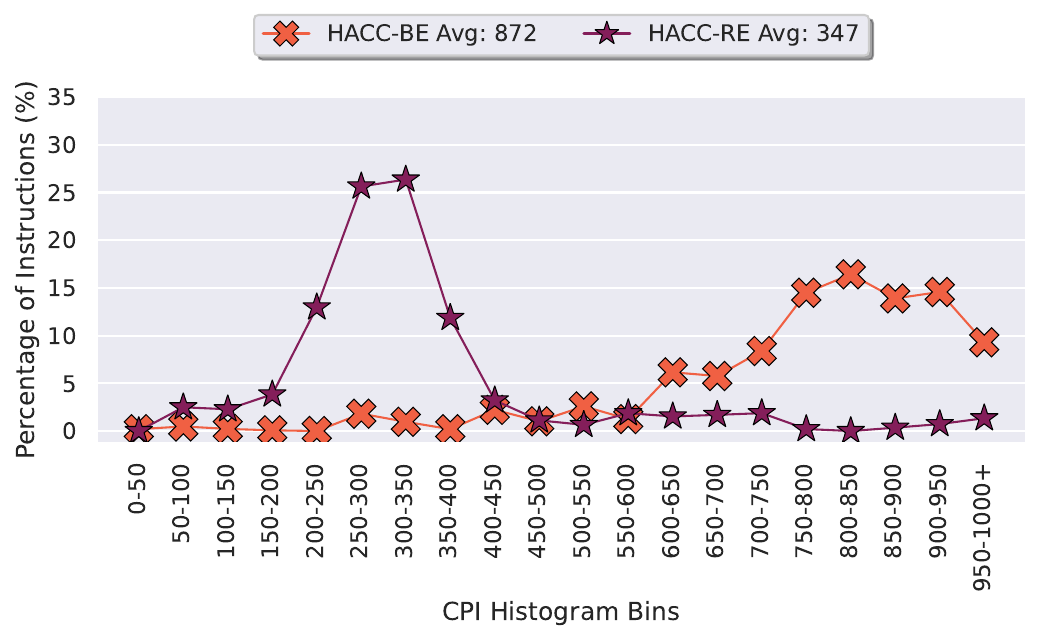}
	\vspace{-1.2em}
	\caption{CPI histogram for \hacc instructions: barrier-based evictions \texttt{HACC-BE} and rolling evictions \texttt{HACC-RE}.}
	\label{fig:hacc_cpi}
\vspace{-1.7em}
\end{figure}

Additionally, the performance of NeuraChip is compared with a row-wise product-based SpGEMM accelerator, Gamma~\cite{zhang2021gamma}, which is based on Gustavson's algorithm. Gamma employs a resource-intensive storage mechanism, FiberCache, to prefetch data, aiming to reduce data fetch latency and prevent compute stalls. However, this approach results in data remaining idle in the caches prior to being accessed by the processing elements. NeuraChip, in contrast, optimizes on-chip storage through a rolling-eviction strategy, enabling automatic eviction of partial products after the reduce operation is complete. Against Gamma, NeuraChip demonstrates a performance superiority of $1.5\times$ average speedup.

\subsection{Comparative Analysis of GNN Accelerators}

\begin{figure}[bp]
	\centering
 \vspace{-1.6em}
	\includegraphics[width=0.45\textwidth]{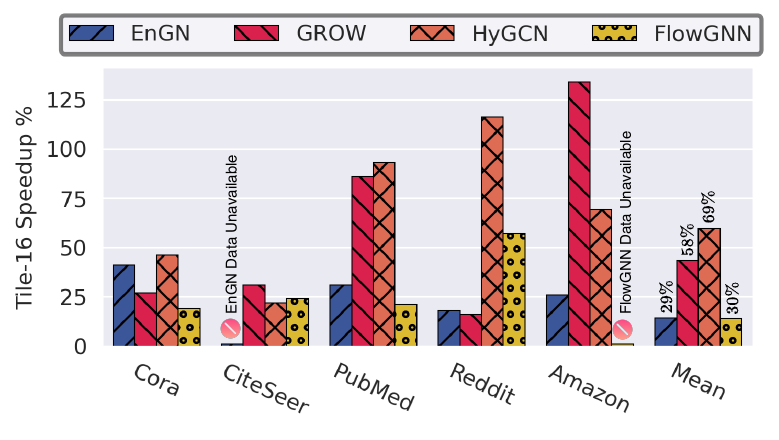}
	\vspace{-1.6em}
	\caption{Tile-16 speedup over previous GNN accelerators for GCN workloads on various graph datasets.}
	\label{fig:gnn_speedup}
\vspace{-1.7em}
\end{figure}

In \figref{fig:gnn_speedup}, we compare the GNN performance of NeuraChip against various state-of-the-art GNN accelerators.
The NeuraChip configuration used for GNN assessment differs from that used to compare to SpGEMM accelerators in Table~\ref{tab:chip_configuration}.
Specifically, for the Tile-16 configuration in the GNN accelerator analysis, an architecture comprising $8$ tiles is used. Each tile includes a $16\times16$ grid of NeuraCores, with each core featuring a quad-pipeline design. We have significantly reduced the number of TAG comparators and port buffers, while retaining the hashpad sizes. This particular configuration is capable of delivering a peak performance of $8192\ \mathrm{GFLOPs}$, with an average power consumption of $4.3$W.

First, we consider EnGN, a hash-based GNN accelerator~\cite{liang2020engn}, and GROW~\cite{hwang2023grow}. EnGN employs a unique ring-based edge reducer to efficiently map vertex IDs. However, it encounters challenges in achieving a uniform distribution of computational tasks among its processing elements. In comparison, NeuraChip demonstrates superior performance, outperforming EnGN by $29\%$ on average. This improvement is primarily attributed to the dynamic reseed hashing function within NeuraChip, which ensures balanced task distribution across its computational resources, namely NeuraCore and NeuraMem, thus minimizing processing delays.

GROW utilizes a row-wise multiplication method, incorporating hardware and software co-design elements. A notable aspect of GROW’s software strategy is its reliance on graph partitioning, which significantly increases the computational overhead for GNN processing. From a hardware perspective, GROW is equipped with vector processors and employs streaming buffers for handling input and output matrix data. Despite these features, GROW encounters issues similar to those seen in Gamma's prefetcher system, where data idling results in suboptimal usage of on-chip memory resources. Comparative performance metrics indicate that NeuraChip surpasses GROW’s performance by an average of $58\%$.

\input{tables/area}

\input{tables/spgemm_compare}

Next, we evaluate our accelerator compared to HyGCN, a hybrid Graph Neural Network (GNN) accelerator, which has specialized engines for aggregation and combination phases~\cite{yan2020hygcn}. The primary advantage of HyGCN's architecture is its ability to pipeline computations, which is particularly beneficial for GNN layers that typically alternate between aggregation and combination phases. However, a significant limitation arises when the compute duration for one phase substantially exceeds the other, leading to a pipeline stall due to the uneven execution duration of each pipeline stage.

Instead, NeuraChip incorporates distinct components specifically for multiplication and accumulation operations, utilized in both the aggregation and combination phases. This design choice renders NeuraChip impervious to the inefficiencies caused by varying computational times between aggregation and combination phases. On average, NeuraChip outperforms HyGCN's performance by $69\%$.

Our final comparison is with FlowGNN~\cite{sarkar2023flowgnn}, a reconfigurable dataflow GNN accelerator comprising Node Transformation Units (NTs) and Message Passing Units (MPs). FlowGNN uses queues for real-time task buffering and relies on dynamic pull-based mapping for task distribution to NTs and MPs. In contrast, NeuraChip adopts a push-based mapping strategy for multiplication tasks and a hash-based approach for accumulation. The Dispatcher in NeuraChip assigns \mmhf instructions to NeuraCores, optimizing input data temporal locality (reuse in NeuraCore register files). The dynamic reseeding hash-based mapping, as detailed in \secref{ssec:drhm}, ensures uniform workload distribution regardless of sparsity patterns. Consequently, NeuraChip achieves an average speedup of $30\%$ over GCN workloads tested on the FlowGNN architecture.

\subsection{Power Consumption and Area Analysis}
\vspace{-0.3em}

We assess our accelerator's area and power overheads by implementing its design in Register Transfer Level (RTL). Using Cadence Genus Synthesis Solutions, we synthesize these RTL components targeting an ASAP7 technology library~\cite{clark2016asap7}, allowing us to determine the area and power consumption for each proposed microarchitectural element. The synthesized chip area requirements for NeuraChip amount to $2.37mm^2$, $10.2mm^2$, and $35.26mm^2$ for the Tile-$4$, Tile-$16$, and Tile-$64$ configurations, respectively.
The breakdown of NeuraChip's area and power is shown in Table~\ref{tab:power}. The majority of the area requirement for NeuraChip is allocated to the NeuraMem unit, as it includes the tag comparator array and the hash-pad (on-chip storage).





%% file: tables/area.tex
\begin{table}
\centering
\caption{NeuraChip Power and Area Breakdown for SpGEMM workloads}
\vspace*{-2mm}
\label{tab:power}
\resizebox{1\linewidth}{!}{%
\begin{tabular}{p{0.55in} | x{0.4in} x{0.4in} x{0.4in} | x{0.4in} x{0.4in} x{0.4in}}
\arrayrulecolor{black}\toprule
     & \multicolumn{3}{c|}{Area ($mm^2$)} & \multicolumn{3}{c}{Average Power (W)}  \\
    \textbf{Unit} &   \textbf{Tile-4} & \textbf{Tile-16} & \textbf{Tile-64} &   \textbf{Tile-4} & \textbf{Tile-16} & \textbf{Tile-64} \\ [0.5ex] 
\arrayrulecolor{black}\toprule
\arrayrulecolor{black}\toprule
      NeuraCore & 0.28 & 2.74 & 9.36 & 1.05 & 1.86 & 5.76  \\
    \arrayrulecolor{black!20}\midrule
        NeuraMem & 1.22 & 5.10 & 18.64 & 6.85 & 7.36 & 11.19  \\
    \arrayrulecolor{black!20}\midrule
        Router & 0.49 & 1.98 & 6.88 & 2.15 & 4.88 & 4.43  \\
    \arrayrulecolor{black!20}\midrule
        Memory Controller & 0.38 & 0.38 & 0.38 & 1.41 & 1.96 & 2.84  \\
    \arrayrulecolor{black}\toprule
        \textbf{Total} & 2.37 & 10.2 & 35.26 & 11.46 & 16.06 & 24.22  \\

     \arrayrulecolor{black}\bottomrule
\end{tabular}
}
\vspace{-6mm}
\end{table}

%% file: tables/spgemm_compare.tex
\begin{table*}[t]
\centering
\caption{Performance comparison of state-of-the-art SpGEMM accelerators across various NeuraChip system configurations.}
\vspace*{-4mm}
\label{tab:spgemm_compare}

\begin{tabular}{x{0.90in} | x{0.40in} x{0.47in} x{0.40in} x{0.40in} x{0.61in} x{0.45in} x{0.7in} x{0.7in} x{0.7in} } 
    \arrayrulecolor{black}\toprule
   \textbf{Architectural Parameters} & Xeon E5 & NVIDIA H100 & AMD MI100 & Outer SPACE & SpArch & Gamma & \textbf{NeuraChip Tile-4} & \textbf{NeuraChip Tile-16} & \textbf{NeuraChip Tile-64} \\
   \arrayrulecolor{black}\toprule
   \arrayrulecolor{black}\toprule
   Compute Units & 8 Cores AVX2 & 7296 FP64 & 7680 FP64 & $256$~PEs & $2\times8$~\small{Mults} $16~\times~16$~\small{Merger}  & $32$~PEs Radix-$64$ & $2\times4$ NeuraCores & $2\times16$ NeuraCores & $2\times64$ NeuraCores \\
   \arrayrulecolor{black!20}\midrule
   Frequency $(\mathrm{GHz})$ & $2.9$ & $1.6$ & $1.5$ & $1.5$ & $1$ & $1$ & $1$ & $1$ & $1$ \\
   \arrayrulecolor{black!20}\midrule
   Peak Performance & $186$ $\mathrm{GFLOPs}$ & $26$ $\mathrm{TFLOPs}$ & $11.5$ $\mathrm{TFLOPs}$ & $384$ $\mathrm{GFLOPs}$ & $32$\ \ \ \  $\mathrm{GFLOPs}$ & $32$ $\mathrm{GFLOPs}$ & $8$\ \ \ \ \ \ \ \ \ \  $\mathrm{GFLOPs}$ & $32$\ \ \ \ \ \ \  $\mathrm{GFLOPs}$ & $128$\ \ \ \ \ \  $\mathrm{GFLOPs}$ \\
   \arrayrulecolor{black!20}\midrule
   SpGEMM Perf.$^{\Phi}$ $(\mathrm{GOP}/s)$ & $1.12$ & $1.86$ & $1.48$ & $2.9$ & $10.4$ & $16.5$ & $5.15$ & $24.75$ & $30.69$\ \ \ \ \  $93.17^{\alpha}$  \\
   \arrayrulecolor{black!20}\midrule
   On-chip Memory & $15~\mathrm{MB}^{\tau}$ & $50~\mathrm{MB}^{\dagger}$ & $8~\mathrm{MB}^{\dagger}$ & $4~\mathrm{MB}$ & $15~\mathrm{MB}^{\star}$ & $3~\mathrm{MB}^{\ast}$ & $0.75~\mathrm{MB}^{\delta}$ & $3~\mathrm{MB}^{\delta}$ & $12~\mathrm{MB}^{\delta}$ \\
   \arrayrulecolor{black!20}\midrule
   Off-chip Memory & DDR4 $136\mathrm{GB}/\mathrm{s}$ & HBM $2\mathrm{TB}/\mathrm{s}$ & HBM $1.2\mathrm{TB}/\mathrm{s}$ & HBM $128\mathrm{GB}/\mathrm{s}$ & HBM $128\mathrm{GB}/\mathrm{s}$ & HBM $128\mathrm{GB}/\mathrm{s}$ & HBM $128\mathrm{GB}/\mathrm{s}$ & HBM $128\mathrm{GB}/\mathrm{s}$ & HBM $128\mathrm{GB}/\mathrm{s}$ \\
   \arrayrulecolor{black!20}\midrule
   Technology $(nm)$ & $32$ & $4$ & $7$ & $32$ & $40$ & $45$ & $7$ & $7$ & $7$ \\
   \arrayrulecolor{black!20}\midrule
   Area $(mm^2)$ & $356$ & $814$ & $750$ & $86.74$ & $28.49$ & $30.6^{\ddagger}$ & $2.37$ & $10.2$ & $35.26$ \\
   \arrayrulecolor{black!20}\midrule
   Power $(W)$ & $85^{\diamond}$ & $300^{\diamond}$ & $300^{\diamond}$ & $24$ & $9.26$ & \ding{118} & $11.46$ & $16.06$ & $24.22$ \\
   \arrayrulecolor{black!20}\midrule
   Energy Efficiency $(\mathrm{GOPS}/W)$ & \ding{117} & \ding{117} & \ding{117} & $0.120$ & $1.123$ & \ding{118} & $0.449$ & $1.541$ & $1.266$ \\
   \arrayrulecolor{black!20}\midrule
   Area Efficiency $(\mathrm{GOPS}/mm^2)$ & \ding{117} & \ding{117} & \ding{117} & $0.034$ & $0.365$ & $0.539$ & $2.171$ & $2.426$ & $0.870$ \\
   \arrayrulecolor{black!20}\midrule
   \textbf{Tile-16 Speedup} & $22.1\times$ & $13.3\times$ & $16.7\times$ & $6.6\times$ & $2.4\times$ & $1.5\times$ & $4.8\times$ & $1\times$ & $0.807\times$ \\
   \arrayrulecolor{black}\bottomrule

\end{tabular}

\vspace*{0.5mm}

\scriptsize


$\diamond$Max thermal dissipation power from datasheet
\ding{118}Gamma lacks a power performance model
$^{\dagger}$ L2 cache size
$^{\tau}$ L3 cache size
$^{\delta}$HashPad Size
$^{\ast}$FiberCache Size
\ding{117}Power and area metrics sourced from vendor datasheets; derived metrics excluded.
$^{\alpha}$ Simulated with dual stacked HBM offering peak bandwidth of $256$ GB/s.
${\ddagger}$Gamma synthesizes accelerator using $45$ nm and $40$ nm processes, resulting in computing areas of $30.6$ $mm^2$ and $20.44$ $mm^2$, respectively.
$^{\star}$Represents column fetchers, row prefetchers, and partial matrix fetchers and writers.
$^{\Phi}$Computed on common set of matrices as shown in Table~\ref{tab:bloat}.

\vspace*{-2.0em}

\end{table*}

%% file: sections/08_related_work.tex
\vspace{-0.5em}
\section{Related Work}
\label{sec:related_work}


\textbf{SpGEMM Accelerators}: 
The InnerSP~\cite{baek2021innersp} accelerator uses the inner-product method for matrix multiplication. This method offers advantages, eliminating the need for on-chip memory for accumulation. However, it suffers from limited input data reuse of both matrices, leading to performance issues when the sparsity patterns do not align with their task mapping algorithm.
MatRaptor~\cite{srivastava2020matraptor} employs a row-wise multiplication strategy and a round-robin greedy algorithm for allocating input rows to processing elements (PEs). Although this approach enhances input data reuse, it struggles with irregular sparsity patterns. The simplistic round-robin distribution may result in computational hot spots (as elaborated in \secref{sec:dse}).
SIGMA~\cite{qin2020sigma} offers an SpGEMM accelerator equipped with adaptable interconnects. Utilizing a smart global controller, SIGMA dynamically assigns each non-zero pair to PEs via a Benes network. Despite its efficiency in general SpGEMM tasks, SIGMA is less effective with large sparse matrix computations due to the substantial overhead introduced by its bitmap compression format.

\textbf{GNN Accelerators}: 
LISA~\cite{li2022lisa} performs GNN computations on Coarse-Grained Reconfigurable Arrays (CGRAs). LISA generates a dataflow graph and utilizes a simulated annealing method for mapping. 
I-GCN~\cite{geng2021gcn} aims to enhance data locality through an {\em islandization} strategy, clustering densely connected nodes to reduce off-chip memory accesses. 
However, both the simulated annealing and graph clustering methods introduce considerable computational overheads.

%% file: sections/09_conclusion.tex
\vspace{-0.8em}
\section{Conclusion}



NeuraChip demonstrates the potential advantages that sparse matrix multiplication workloads can gain from a decoupled architectural design.
We have presented an open-source, cycle-accurate simulator called NeuraSim, used to demonstrate the effectiveness of our design. 
GNN workloads acceleration is achieved through a blend of high-level optimizations and microarchitectural features.
We synthesized our design in RTL, giving us power and area requirements for various NeuraChip Tile sizes.  NeuraChip outperforms state-of-the-art SpGEMM accelerator by a factor of $1.5\times$ and prior GNN accelerators by $1.46\times$ on average.
In future work, we plan to explore the fabrication of NeuraChip to fully demonstrate the benefits of our approach.

%% file: sections/99_artifact.tex
\appendix

\section{Appendix}
\subsection{Artifact Evaluation}

NeuraChip is a GNN accelerator that is built using the NeuraSim simulator. NeuraSim is a cycle-accurate simulator that contains six modules as follows:

\begin{enumerate}
    \item \textbf{NeuraSim Engine}: The cycle accurate simulator engine built in C++.
    \item \textbf{NeuraCompiler}: A pythonic compiler that takes graphs as inputs and generates a SpGEMM workload binary to be executed by the NeuraSim Engine. The NeuraCompiler uses an extended $\mathrm{x}86$ ISA.
    \item \textbf{Mongo}: The Mongo module is a part of NeuraSim Engine that is coupled with an instance of MongoDB server. Throughout the simulation, NeuraSim pushes performance metrics to the MongoDB server.
    \item \textbf{NeuraViz}: A pythonic module that generates all plots to visualize the data. NeuraViz is coupled with MongoDB and fetches simulation metrics from the MongoDB server.
    \item \textbf{Dashboard}: For ease of use, NeuraChip is hosted on our servers and accessible to users via a WebGUI named Dashboard, available at \url{https://neurachip.us}.
    \item \textbf{DRAMSim3}: NeuraChip utilizes the memory simulator provided by Li et al., called DRAMSim3~\cite{li2020dramsim3} for computing HBM memory request latencies.
\end{enumerate}

This appendix describes where to access our code artifact and how to reproduce part of our experiments and results. We use a Git repository and host an instance of our simulator, which is available through a public WebGUI, to schedule simulations of different configurations and display results. Additionally, we provide instructions on how to natively build NeuraSim in the project's~\href{https://github.com/NeuraChip/neurachip/blob/main/README.md}{README}.

\subsection{Artifact Checklist}
\begin{itemize}
    \item Benchmarks: SpGEMM workload over a common set of matrices mentioned in Table~\ref{tab:bloat}.
    \item Runtime environment: Tested on Ubuntu 22.04, though it should be reproducible when run on any operating system as long as Docker is utilized. Simulations using the hosted WebGUI only require a web browser.
    \item Hardware: To compile and run the simulation experiments, we recommend a machine with 32 GB of DRAM memory, 50 GB of disk storage, and 8 cores.
    \item Execution: We provide a web interface that allows the execution of different experiments. Docker instructions are also provided so users can build the environment themselves.
    \item Metrics: SpGEMM performance in GFLOP/s, Execution time, NeuraCore utilization, NeuraMem utilization, Histogram of number of cycles spent by instructions in Register File, NeuraRouter Utilization, Average In-Flight Memory Transactions Per Cycle (Memory Pressure).
    \item Output: Experiments deployed on the WebGUI generate simulation metrics, which are used to generate the graphs/plots.
    \item Disk space required: Around 50GB, which includes installing necessary dependencies, building NeuraSim, and running one simulation.
    \item How long does it take to prepare the workflow?: The compilation/installation using Docker takes around 30 minutes on a 32-cores machine with 64GB DDR4 memory.
    \item Experiment completion time: 30 minutes.
    \item Publicly available: Yes,~\url{https://github.com/NeuraChip/neurachip}.
    \item Archived: Yes,~\url{https://doi.org/10.5281/zenodo.10896280}.

\end{itemize}

\subsection{How to access?}



The easiest way to access NeuraChip is via the WebGUI; however, compiling it from source is also possible.
Here are the simplified steps to run simulations using the WebUI:

\begin{enumerate}
    \item Visit \url{https://neurachip.us}
    \item Click on "Launch a new simulation".
    \item On the \textit{New NeuraSim Simulation} page, click on "Launch" and wait for 20 minutes.
    \item Once all the plots are generated, a "Results" button will pop up at the bottom of the page. Click on the "Results" button to view many simulation performance metrics on the \textit{NeuraViz Results} page.
\end{enumerate}

To compile the NauraSim locally on a platform with Docker:

\begin{enumerate}
    \item Clone the repository from GitHub \\
    \texttt{git clone \url{https://github.com/NeuraChip/neurachip.git}}
    \item Enter the directory \\
    \texttt{cd neurachip}
    \item Build Docker image from docker file \\
    \texttt{./build-docker.sh}
    \item Start the docker container \\
    \texttt{./start-docker.sh}
    \item Run MongoDB server within Docker container \\
    \texttt{sudo -u mongodb mongod --config /etc/mongod.conf --fork}
    \item Enter NeuraSim directory \\
    \texttt{cd /home/ktb/git/NeuraChip/NeuraSim}
    \item Compile NeuraSim with compile script \\
    \texttt{./compile-run.sh}
    \item Run the Simulator \\
    \texttt{./chippy.bin}
    \item After execution, the NeuraSim will print to the terminal the total number of cycles for a NeuraChip to execute the workload, duration of simulation, and number of simulated instructions per second. Other metrics are stored in the simulation database.
\end{enumerate}

\subsubsection{Hardware dependencies}: We recommend a minimum of 64 GB of DRAM memory and 50 GB of storage space.

\subsubsection{Software dependencies}

In our execution environment, all software dependencies necessary for operation are fully satisfied.
Additionally, when compiling NeuraSim simulator from scratch, we utilize Docker to manage and meet all the required dependencies.

\subsubsection{Datasets}

Datasets for our simulations are derived from Table~\ref{tab:bloat}. Using the default configurations, we perform simulations of SpGEMM computations on NeuraChip configured with 16 NeuraCores and 16 NeuraMems. These simulations utilize the Cora workload under a setup referred to as the Tile-16 configuration.

\newpage
\subsection{Execution}

We grant evaluators remote access to our simulator's web interface. Below are the steps to interact with the NeuraSim simulator available on our website.

\begin{enumerate}
    \item Accessing the NeuraSim Simulator WeGUI:
    \begin{itemize}
        \item Open your web browser and navigate to the URL provided for the NeuraSim simulator website: \url{https://neurachip.us}.
    \end{itemize}
    
    \item Navigating to Configuration Selection:
    \begin{itemize}
        \item Upon loading the website, you will be directed to the main page of the NeuraSim simulator.
        \item Look for a button labeled \textit{Launch a new Simulation}.
    \end{itemize}
    
    \item Selecting Configurations:
    \begin{itemize}
        \item Once on the Configuration Selection page, you will see a list of selectable presets that configure NeuraChip parameters.
        \item Select the configurations to evaluate.
        \item If you wish to save the results to view later, change the "Experiment Name" to something unique, and after the results have been generated, bookmark the webpage in your browser. To preserve anonymity, do not use personal/recognizable experiment names.
    \end{itemize}
    
    \item Initiating the Simulation:
    \begin{itemize}
        \item After selecting the desired configurations, locate the \textit{Launch} button or option to initiate the simulation process.
        \item Click on the button to start the simulation with the chosen configurations.
    \end{itemize}
    
    \item Monitoring Simulation Progress:
    \begin{itemize}
        \item While the simulation is running, you may observe progress indicators or status updates to track the simulation's progress.
        \item Depending on the complexity of the simulation and the chosen configurations, this process may take some time.
    \end{itemize}
    
    \item Viewing Simulation Results:
    \begin{itemize}
        \item Once the simulation completes, the website should display the \textit{Results} button.
        \item Look for a section labeled \textit{Simulation Results}, where you can observe the outcome of the simulation.
        \item If you have given your experiment a unique \textit{Experiment Name}, you can bookmark this link and re-visit it at a later time. Experiments with unique names are preserved on the database and not deleted even if the browser window is closed.
    \end{itemize}
    
    \item Analyzing the Results:
    \begin{itemize}
        \item Simulation metrics, along with their description, are provided on the web page.
    \end{itemize}
    
\end{enumerate}

%% file: sections/acknowledgement.tex
\begin{acks}
This research was supported in part by the Institute for Experiential AI and the NSF IUCRC Center for Hardware and Embedded Systems Security and Trust (CHEST), NSF CNS 2312275, NSF CNS 2312276, Samsung Advanced Institute of Technology, Samsung Electronics Co., Ltd., NRF-2023R1A2C2004229, and IITP No.RS-2023-00228255. Additionally, we acknowledge the financial assistance from grant RYC2021-031966-I funded by MCIN/AEI/10.13039/501100011033 and the ``European Union NextGenerationEU/PRTR.'', and grant PID2022-136315OB-I00 funded by MCIN/AEI/10.13039/501100011033/ and by ``ERDF A way of making Europe’’, EU.
\end{acks}